\documentclass[aps,pra,reprint,showpacs,twocolumn]{revtex4}
\usepackage{bm}
\usepackage{amsmath, amsfonts,amssymb,mathrsfs}
\usepackage{physics}
\usepackage{color}
\usepackage{soul}
\usepackage{graphicx}
\usepackage[
  colorlinks=true,
  urlcolor=blue,
  linkcolor=blue,
  citecolor=blue
]{hyperref}

\usepackage{epsf,epsfig,graphics,graphicx}
\usepackage{verbatim,color,ulem}
\usepackage{setspace}
\usepackage{graphicx,wrapfig}
\usepackage{enumerate}
\usepackage{color}
\usepackage[ddmmyyyy]{datetime}
\usepackage{mathrsfs}
\usepackage{subfigure}
\usepackage{amssymb,amsmath,amsthm}
\usepackage[mathscr]{eucal}

\begin{document}

\title{Regular and chaotic behavior of collective atomic motion in two-component
Bose-Einstein condensates}
\date{\today}
\author{Wei-Can Syu}
\email{syuweican@gmail.com}
\affiliation{Department of Physics,
National Dong-Hwa University, Hualien, Taiwan, R.O.C.}
\author{Da-Shin Lee}
\email{dslee@gms.ndhu.edu.tw}
\affiliation{Department of Physics,
National Dong-Hwa University, Hualien, Taiwan, R.O.C.}
\author{Chi-Yong Lin}
\email{lcyong@gms.ndhu.edu.tw}
\affiliation{Department of
Physics, National Dong-Hwa University, Hualien, Taiwan, R.O.C.}

\begin{abstract}
We theoretically study binary Bose-Einstein condensates trapped in a
single-well harmonic potential to probe the dynamics of collective atomic motion.
The idea is to choose tunable scattering lengths through Feshbach resonances such that the ground-state wave function for two types of the condensates are spatially immiscible where one  of the condensates, located at the center of the potential trap, can be effectively treated as a potential barrier between  bilateral condensates of the second type of atoms.
In the case of small wave function overlap between bilateral condensates,
 one can parametrize their spatial part of the wave functions in the two-mode approximation together with the time-dependent population imbalance $z$ and the phase difference $\phi$ between two wave functions.
The  condensate in the middle  can be approximated by a Gaussian wave function with the displacement of the condensate center $\xi$.
As driven by the time-dependent displacement of the central condensate, we find the Josephson oscillations of the collective atomic motion between bilateral condensates as well as their anharmonic generalization of  macroscopic self-trapping  effects.
In addition, with the increase in the wave function overlap of bilateral condensates by properly choosing tunable atomic scattering lengths, the chaotic oscillations are found if the system departs from the state of a fixed point. The Melnikov  approach with a homoclinic solution of the derived  $z,\,\phi$, and $\xi$  equations can successfully justify the existence of chaos.
All results are  consistent with the numerical solutions of the full time-dependent Gross-Pitaevskii equations.
\end{abstract}


\maketitle

\section{Introduction}\label{sec1}
 The first experimental observation of interference fringes between two freely expanding Bose-Einstein condensates (BECs)
has launched the fascinating possibility to probe new quantum phenomena on macroscopic scales related to the superfluid nature of the Bose condensates  \cite{Andrews1997}.
A chief effect is the Josepshon analog, in which the phase differences between two trapped BECs in a double-well potential can generate Josephson-like current-phase oscillations via collectively  atomic tunneling through the barrier \cite{Smerzi1997,Raghavan1999}.
Nevertheless, the nonlinearity of tunneling effects produces novel
anharmonic Josephson oscillations and macroscopically quantum
self-trapping (MQST) effects.
The potentially existing such interesting phenomena indeed triggers a wide variety of theoretic investigations in the settings of two-component BECs systems in the double-well potential \cite{JuliaDiaz2009,Sun2009,Satija2009,Messeguer2011},   triple-well \cite{Richaud2018} and
 the single BEC system in the triple-well \cite{Liu2007}, four-well \cite{Liberato2006} and multiple-well potentials \cite{Cataliotti2001,Nigro2018}, to cite a few. The effective Josephson dynamics can also be observed in some
 systems without an external potential~\cite{aba}.
More importantly, several experiments have successfully observed  coherence oscillations through quantum tunneling of  atoms  between the condensates trapped in a double-well potential \cite{Albiez2005,LeBlanc2011,Pigneur2018}, whereas similar phenomena were also seen from attractive to repulsive atomic interactions by experimentally changing atomic scattering lengths through  Feshbach resonances \cite{Spagnolli2017}.
In addition to the regular coherent oscillations mentioned above, the studies on the possibility of the appearance of chaotic motions,  by applying  an externally time-dependent trap potential, are also a fascinating task with findings that deserve further experimental justification
\cite{Abdullaev2000,Lee2001,Jiang2014,Tomkovic2017}. Chaotic behavior is also studied in the dynamics of of three coupled BECs~\cite{fra}.

In the present work, we consider the two-component BECs system with atoms
  in two different hyperfine states
 in a single-well harmonic trap potential. Such a two-component BEC system provides us an ideal platform for the study of intriguing phenomena, for example, mimicking quantum gravitational effects in Ref.~\cite{Syu2019}.
Using Feshbach resonances to experimentally tune the scattering lengths of atoms
allows us to construct the diagram in Fig.~\ref{fig:phasediagram0}, showing the typical ground-state wave function of the binary condensates \cite{Thalhammer2008,Tojo2010,Catani2008,Wacker2015,Moses2015}.
The effective parameter characterizing the miscibility or immiscible regime of binary condensates is mainly determined by the value of $\Delta = g_{11}g_{22}-g_{12}^2$, where $g_{ij}$ denotes the atomic interaction between $i$ and $j$ atoms \cite{Riboli2002,Papp2008,Sasaki2009}.
The parameter regime for $\Delta>0$ corresponds to the  miscible distribution of two-component condensates, whereas for $\Delta<0$ the different species of atoms repel each other and
the distribution becomes immiscible \cite{Riboli2002,Papp2008,Sasaki2009}.
In particular, we focus on the situations of the scattering lengths in the immiscible regime, where the condensate of one of the components is located at the center of the potential trap, and can effectively be regarded as an effective potential barrier between bilateral condensates of the second component of the system.
 Experimentally, spatially asymmetric bilateral condensates can be prepared by means of a magnetic field gradient \cite{McCarron2011, Eto2015, Eto2016}, allowing the possibility to observe collectively coherence oscillations of  atoms between them.
 This is an extension of the works of \cite{Smerzi1997,Raghavan1999}, where
 the system of a single BEC in an external double-well potential is studied. The collective oscillations of  atoms through  quantum tunneling between the condensates centered at each of the potential wells have been observed where
 the full dynamics of the system can be further simplified, and turned  into an analog pendulum's dynamics  in the so-called two modes approximation \cite{Burchianti2017,Smerzi1997,Raghavan1999,Ananikian2006}.
%
In our setting, the dynamics of the system can also be approximated by  analogous coupled pendulums dynamics using the variational approach as well as two modes approximation. One can then reproduce regular oscillations around the stable states studied in Refs. \cite{Smerzi1997,Raghavan1999} within some parameter regime by considering the effects from time-averaged trajectories of the  central condensate.
Additionally, the collective motion of atoms, if departing from the unstable states, could show chaotical behavior, oscillating between bilateral condensates due to the time-dependent driving force given by the dynamics of the  condensate centered in the middle  \cite{Abdullaev2000,Lee2001,Jiang2014,Tomkovic2017}. The scattering lengths in the parameter regime, which give the chaotic dynamics of the system, will be useful as a reference for experimentalists to observe the effects.

To be concrete, consider  a mixture of $^{85}\text{Rb}$  $\vert 1\rangle=\vert F=2, m_F=-2 \rangle$
and $^{87}\text{Rb}$  $\vert 2 \rangle=\vert F=1, m_F=-1\rangle$ by ignoring the small mass difference~\cite{Papp2008}.  The
scattering lengths
 we choose are estimated to be $a_{22}= 50 a_0$ and $a_{12}= 85 a_0$ with $a_0$ the Bohr radius.
The existence of the Feshbach
resonance at $160 \text{G}$ permits us to tune the  scattering length $a_{11}$ of $^{85}\text{Rb}$  atoms from
$50a_{0}$ to $150a_{0}$ to be specified in each case later \cite{Papp2008}.
Let us consider two-component elongated BECs in quasi-one-dimensional (1D)  geometries with  the trap potential {\cite{Becker2008}  $\omega_{x}=2\pi\times
12\text{Hz},\,\omega_{y,z}=2\pi\times 100\text{Hz}$.
We take the numbers of atoms $N_1=500$, and $N_2=1000$ as an example.
The experimental realization  of the Josephson dynamics  by two weakly linked BECS in a double-well potential
in a single BEC system is confirmed in Ref.~\cite{Albiez2005}. Both Josephson oscillations and nonlinear trapping are observed by tuning the
external potential. Such phenomena are also explored in two-component BECs system in a double-well potential where apart from
their own atomic scattering processes the Rabi coupling between the atoms from each components of the BECs is introduced. Tuning
the scattering lengths using the associated Feshbach resonances and/or the Rabi coupling constant allow us to observe the
transition from Josephson oscillations to nonlinear self-trappings~\cite{zib}.
 Our proposal also in the two-component BECs system but in a single-well potential suggests a novel setup, although involving
 more rich dynamical aspects of  condensates, the only tunable parameters are  atomic scattering lengths again through Feshbach
 resonances to achieve the miscible condensates that needs more experimental endeavor as compared with~\cite{zib}. 

\begin{figure}
\begin{center}
\includegraphics[width=1\linewidth]{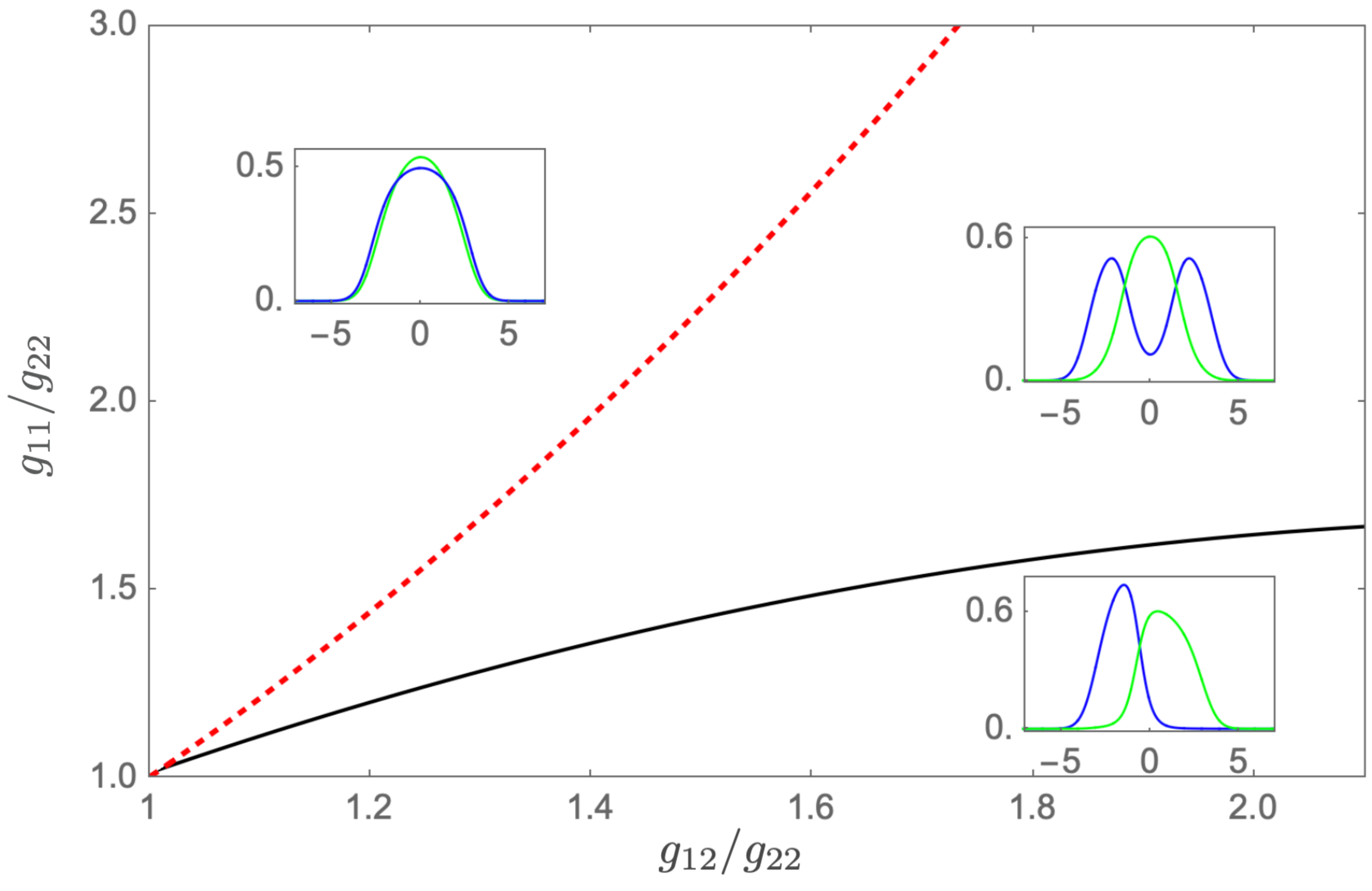}
\caption{The ground-state wave function of binary BECs system in a cigar-shape trap potential is shown for  two species $^{85}\text{Rb}$ $|2,-2\rangle$ state and $^{87}\text{Rb}\,|1,-1\rangle$ state  with atomic coupling constants in miscible/immiscible regimes respectively.
The numbers of atoms, say  $N_1=500,\,N_2=1000$, and also  the   potential trap  $\omega_x=2\pi\times 12\text{Hz},\,\omega_{y,z}=2\pi\times 100\text{Hz}$ are  chosen to show the spatial distribution of wave functions in the immiscible  (miscible) regime for $\Delta <0$ ($\Delta >0$) below (above) red dashed line determined by $\Delta=0$, where $\Delta=g_{11} g_{22}-g_{12}^2$. The black solid line separates between symmetric and asymmetric condensate wave functions.}
\label{fig:phasediagram0}
\end{center}
\end{figure}

Our presentation is organized as follows.
In Sec.~\ref{sec2}, we first introduce the model of the binary BECs system, and the corresponding time-dependent Gross-Pitaeviskii (GP) equations for each of condensate wave functions. In this two-component BECs system, we study the evolution of the wave functions within a strong cigar-shape trap potential so that it can effectively treated as a quasi one-dimensional system.
We further assume that the  condensates start being displaced from their ground state in the
immiscible regime where one of the condensates is distributed in the trap center while the other condensate surrounds the  central condensate  on its two sides.
Then we propose the ansatz of the Gaussian wave functions, which can reduce the dynamics of the system into few degrees of freedom, such as the center of the central condensate as well as the population imbalance and the relative phase difference between bilateral condensates in the two-mode approximation. We later obtain their equations of motion with the form of  coupled oscillators equations.
In Sec.~\ref{sec3}, we analyze the stationary-state solutions and their stability property.
In Sec.~\ref{sec4}, we examine the time evolution of the system around the stable stationary state,  showing the regular oscillatory motion.
In Sec.~\ref{sec5}, the study  is focused on when the system starts from  near unstable stationary-state solutions, exhibiting the chaotic behavior by numerical studies. The Melnikov homoclinic method is adopted for analytically studying the existence of chaos.
Concluding remarks are in Sec.~\ref{sec6}.
In Appendix, we provide more detailed derivations or approximations to arrive at  coupled oscillators equations from  the time-dependent GP equations using the appropriate ansatz of Gaussian wave functions.

\section{Variational approach and {analogous coupled pendulums dynamics} }\label{sec2}

We consider the binary BECs for  same atoms in two different hyperfine states confined in  a strong  cigar-shape  potential with the size of the trap $L_x$ along  the axial direction,taken in the $x$ direction,  which is much larger than  $L_r$ along the radial direction. This system can  be effectively treated as the pseudo one-dimensional system with the Lagrangian described as \cite{Garcia1996,Salasnich2002}
\begin{align}
L_\text{1D}=&\sum_{j=1,2}\int dx\Bigg[\frac{i\hbar}{2}\left(\psi_j\frac{\partial \psi_j^*}{\partial t}-\psi_j^*\frac{\partial \psi_j}{\partial t}\right)\nonumber\\
&\qquad\qquad-\left(\frac{\hbar^2}{2 m}\left| \frac{\partial\psi_j}{\partial x}\right|^2+ V_{\rm ext} |\psi_j|^2+\frac{g_{jj}}{2}|\psi_j|^4\right)\Bigg]\nonumber\\
&\quad+\int dx \, g_{12}|\psi_1|^2|\psi_2|^2,
\label{Lag}
\end{align}
where  the external potential in the axial direction is given by $V_{\rm ext} (x) =m \, \omega_x^2 \, x^2/2$. The mass of  atoms is $m$ and the coupling constants between atoms are given in terms of  scattering lengths $a_{ij}$ as $g_{ij}=2\hbar^2 a_{ij}/m L_r^2$. The wave function  $\psi_j$ is subject to the normalization
\begin{align}
\int d \, {x}\, |\psi_j(\mathbf {x})|^2 =N_j
\end{align}
 with the numbers of atoms $N_j$ in each hyperfine states $j$. The time-dependent GP equations for each component of the BEC condensates are given by \cite{Pethick2008}
 \begin{align}
i\hbar\frac{\partial \psi_1}{\partial t}=\left(-\frac{\hbar^2}{2m}\frac{\partial^2}{\partial x^2}+V_\text{ext}+g_{11}|\psi_1|^2+g_{12}|\psi_2|^2\right)\psi_1,\label{GPe1}\\[7pt]
i\hbar\frac{\partial \psi_2}{\partial t}=\left(-\frac{\hbar^2}{2m}\frac{\partial^2}{\partial x^2}+V_\text{ext}+g_{22}|\psi_2|^2+g_{12}|\psi_1|^2\right)\psi_2.
\label{GPe2}
\end{align}
The full dynamics of the condensates in a harmonic trap potential can be explored by solving the GP equations numerically with given initial wave functions. However, in order to extract their feature analytically, we will propose an ansatz of the wave functions with several relevant parameters.  All results from solving the equations of motion for the introduced parameters  given by the variational approach will be justified by  comparing with those given by numerically solving the GP equations.
Moreover,
we will rescale the spatial or temporal variables into the dimensionless ones by setting $t \rightarrow t/\hbar \omega_x$ and $ x \rightarrow  x/\sqrt{\hbar/m\omega_x}$. Henceforth, the dimensionless $t$ and $x$ will be adopted, and their full dimensional expressions will be recovered, if necessary.

Experimentally, the values of parameters $g_{ij}$, which are related to scattering lengths $a_{ij}$, can be tuned by Feshbach resonances ~\cite{Thalhammer2008,Tojo2010,Papp2008}.
In fact, we might explore
 the evolution of condensate wave functions, in which $\psi_1$ and $\psi_2$
 are initially prepared
 in the immiscible regime,  determined by the  parameter $\Delta = g_{11}g_{22}-g_{12}^2$ when $\Delta<0$  \cite{Riboli2002,Papp2008,Sasaki2009} as illustrated in Fig.~\ref{fig:phasediagram0}.
Let us now assume the general Gaussian wavefunction $\psi_2$ centered at the potential trap center as \cite{Garcia1996,Salasnich2002, Lin2000}
\begin{align}
&\psi_{ 2}(x,t)=\nonumber\\
&\,\,\left(\frac{{N_2}^2}{\pi w(t)^2}\right)^{1/4}\exp[-\frac{(x-\xi(t))^2}{2w(t)^2}+ix\alpha(t)+ix^2\beta(t)],
\label{psi2}
\end{align}
where four time-dependent variational parameters are the  position of the center $\xi(t)$, the width $w(t)$, and $\alpha(t)$ (slope) and $\beta(t)$ $ ({\rm (curvature)}^{1/2}$) of the wave function.
%
We also assume that the ground-state configuration of
$\psi_1(x,t)$
is of the two-mode form \cite{Smerzi1997,Raghavan1999,Burchianti2017,Ananikian2006}
\begin{align}
\psi_{1} (x,t)=\varphi_L(t)\psi_L(x)+\varphi_R(t)\psi_R(x) \, .
\label{psi1}
\end{align}
 \begin{figure}[t]
\begin{center}
\includegraphics[width=0.9\linewidth]{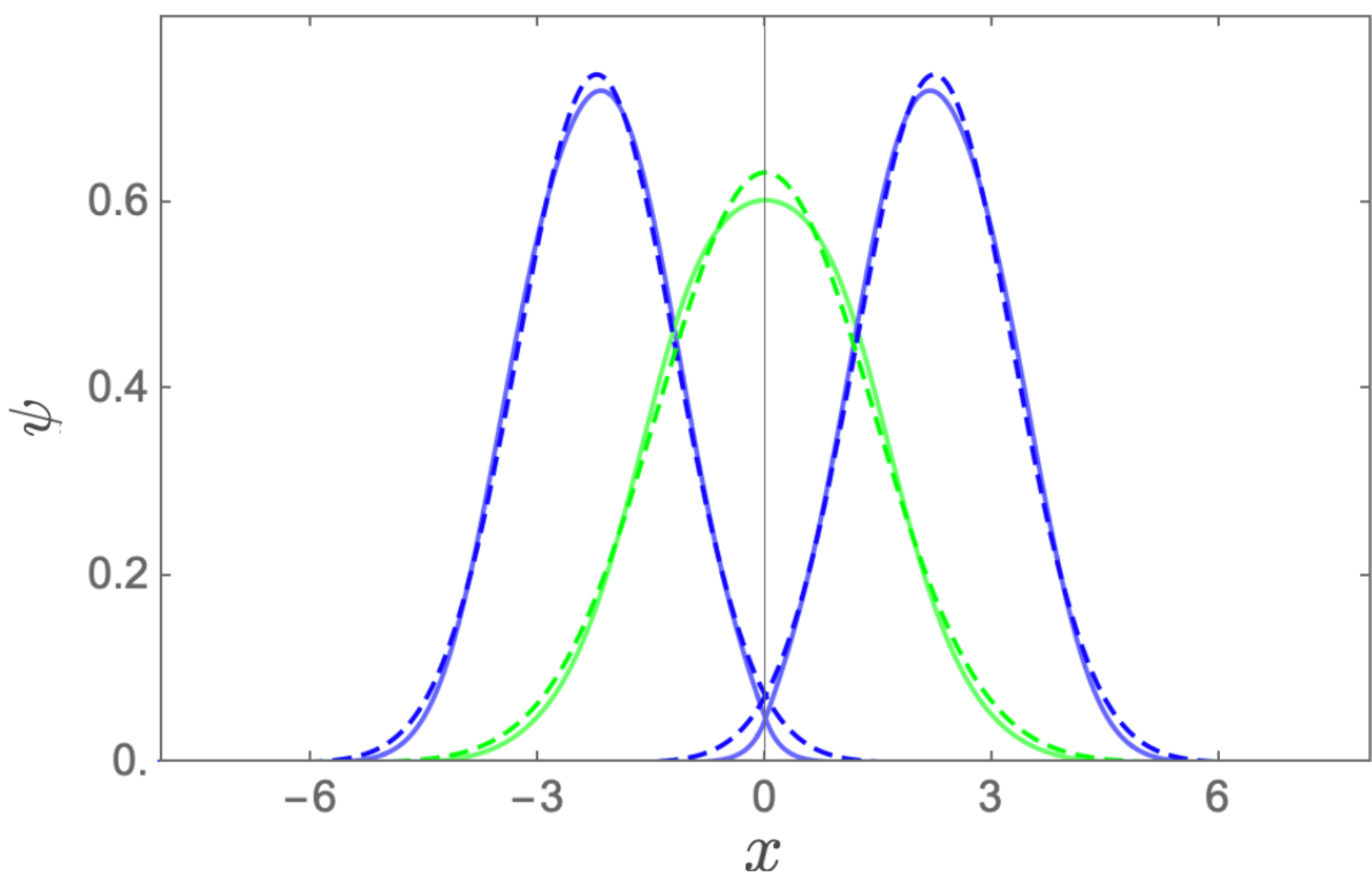}
\caption{The typical condensate wave functions are plotted for the stationary-state solutions of the GP equations
for scattering length $a_{11}=110a_{0}$, $a_{22}=50a_{0}$ and $a_{12}=85a_{0}$ that give  $g_{11}/g_{22}=2.2$ and $g_{12}/g_{22}=1.7$ ($\Delta<0$ in the immiscible regime shown in Fig.~\ref{fig:phasediagram0}), which can be
 fitted into a Gaussian function (dashed line).}
\label{fig:wavefun}
\end{center}
\end{figure}
%
The involved Gaussian-like spatial functions $\psi_L(x)$ and $\psi_R(x)$  together with the initial wave function
$\psi_{1}(x,t=0)$ with width $\sigma_0$ will be determined numerically by the stationary-state solutions
of the  GP equations~(\ref{GPe1}) and (\ref{GPe2}) with the forms shown in Fig.~\ref{fig:wavefun}, which is symmetric
with respect to $x=0$.
$\psi_L(x)$ and $\psi_R(x)$ presumably serve as a good basis to express the spatial part of the wave functions of bilateral condensates with normalization conditions $\int dx \, |\psi_{L,R}|^2=1$, and also negligible overlap $\int dx \,  \psi_L\psi_R \approx 0$ for the weak link between bilateral condensates
 in the  immiscible  regime. Note that the analytical approach based on the wave functions above gives transparent and consistent results as compared with those from numerically solving the full GP equations in the case of Josephson oscillations but for the motion of MQST discrepancy arise as will be seen later where one should look for the wave functions of bilateral condensates being not restricted to be symmetric with respect to $x=0$. 
 Thus, the time-dependent part of the wave function can be parametrized as
 \begin{align}
&\varphi_L(t)=\sqrt{N_L(t)}\, e^{i\phi_L(t)}, \\
&\varphi_R(t)=\sqrt{N_R(t)}\, e^{i\phi_R(t)}
\end{align}
with the phase $\phi_{L,R}(t)$ and the amplitude $\sqrt{N_{L,R} (t)}$ given by the number of atoms in each sides of the
condensates. The total number of atoms ($j=1$) is $N_1=N_L+ N_R$.
To realize quantum coherence between left and right condensates,
and also atomic number oscillations,
it finds more convenient to introduce the  population imbalance
\begin{align}
z(t)\equiv\frac{N_L(t)-N_R(t)}{N_L(t)+N_R(t)},
\end{align}
and relative phase
\begin{align}
\phi(t)\equiv\phi_R(t)-\phi_L(t).
\end{align}
When all atoms locate in the left(right) side of the central condensate, $z=1(-1)$.

The time evolution of coherent oscillations between bilateral condensates surrounding around the central condensate is mainly determined not only by the self-interaction of atoms~\cite{Smerzi1997,Raghavan1999}, but also the interspecies coupling constant $g_{12}$  with the wave functions $\psi_1(x,t)$ and $\psi_2(x,t)$,
which give an additional effect to influence the dynamics of coherent oscillations due to the wave-function overlap between them.
Also, notice that although the initial wave function of the central condensate is centered at $x=0$, which is  the minimum of the external trapped potential, the later evolution of the wave function will find its stationary state with the center located at $x=\xi_0$, moving around that position with nonzero kinetic energy.
Thus, in the presence of $\psi_2(x,t)$  shown in Fig.~\ref{fig:wavefun}
together with the external trapped potential,
the resulting effective potential experienced by the wave function $\psi_1(x,t)$ in the GP equation in (\ref{GPe1}) can be seen in the shape of the double-well potential.
However, the dynamics of the central condensate leads to  the effective potential time dependent, contributing its effects to quantum coherence oscillations between bilateral condensates. This will be the main purpose of studies in Secs. IV and V.
This is an extension of the work in Refs.  \cite{Smerzi1997,Raghavan1999} with additional degrees of freedom of the central condensate. When the initial conditions of the central or bilateral condensates are chosen near their stable-state configurations, the dynamics of coherence oscillations
modifies slightly the behavior found in Refs.  \cite{Smerzi1997,Raghavan1999}.
%
%
Nevertheless, as their initial conditions are chosen near the unstable-state configurations, the chaos occurs. Although the
 analogous coupled pendulums dynamics, which is reduced from the full dynamics of the system and later gives essential information on the time evolution of the condensate wave functions,
 may produce relatively different trajectories from the solutions of the time-dependent GP equations, they are still very useful to provide us analytic analysis on the chaos dynamics by adopting the Melnikov Homoclinic method.

Substituting the ansatz of the wave function (\ref{psi2}) and (\ref{psi1}) into the Lagrangian (\ref{Lag}), and  carrying out the integration over space, the effective action
then becomes the functional of time-dependent variables $\alpha$, $\beta$, $\xi$, $w$, $z$, and $\phi$.
%
Their time evolutions follow the Lagrange equations of motion derived from this effective Lagrangian, discussed in detail in the Appendix.
When
the wave function $\psi_2(x,t=0) $ is slight away from the stationary state,
the dynamics of the condensate just undergoes small amplitude oscillations around that state.
We will further assume that the displacement $\xi$ and the wave function width, deviated from $\sigma_0$ and  defined by $w=\sigma_0+\sigma$, are small as compared with $\sigma_0$, namely $\xi \ll \sigma_0$ and $\sigma \ll \sigma_0 $.
%
%

Retaining terms up to linear order in $\xi$ and $\sigma$ of interest, we have found the Lagrange equations for the  parameters  by the variational approach as
\begin{align}
&\dot z+2\left(k+\kappa  \frac{\sigma}{\sigma_0}\right)\sqrt{1-z^2}\sin\phi=0\label{ap2_z},\\[7pt]
&\dot \phi-\Delta E-\Lambda z \nonumber\\
&\quad -2\left(k+\kappa \frac{\sigma}{\sigma_0}\right)\frac{z}{\sqrt{1-z^2}}\cos\phi+ 2 \sqrt{N_2/N_1} \, \eta \xi=0,\label{ap2_phi}\\
&\ddot{\xi}+\omega_\xi^2\xi-\sqrt{N_1/N_2} \, \eta z-\varepsilon\sigma=0,\label{ap2_xi}\\[7pt]
&\ddot{\sigma}+\omega_\sigma^2\sigma-2\varepsilon\xi=0 \, . \label{ap2_sigma}
\end{align}
The quantities that appear in the equations above are the integrals involving the condensate wave functions, namely
\begin{align}
&k_0=-\int dx\left(\frac{1}{2}\frac{d\psi_L}{dx}\frac{d\psi_R}{dx}+\frac{x^2}{2}\psi_L\psi_R\right), \label{k0}\\
&k=k_0-\frac{g_{12}N_2}{\sqrt{\pi}\sigma_0}\int{dx\psi_L\psi_R\,e^{-x^2/\sigma_0^2}}
\; , \label{k}\\
&\kappa=\frac{g_{12}N_2}{\sqrt{\pi}\sigma_0}\int dx\psi_L\psi_R\left(1-2\frac{x^2}{\sigma_0^2}\right)e^{-x^2/\sigma_0^2}
\; ,\label{kapp} \\
&\Delta E =\int dx\left[\frac{1}{2}\left(\frac{d\psi_L}{dx}\right)^2+\frac{x^2}{2}\psi_L^2+\frac{g_{11}N_1}{2}\psi_L^4\right]\nonumber\\
&\qquad-\int dx\left[\frac{1}{2}\left(\frac{d\psi_R}{dx}\right)^2+\frac{x^2}{2}\psi_R^2+\frac{g_{11}N_1}{2}\psi_R^4\right]\label{DE}
,  \\
&\Lambda \,=\, \frac{g_{11}N_1}{2}\left(\int dx \psi_L^4+\int dx \psi_R^4\right)
\; ,\label{lambda}\\
&\eta= -\frac{g_{12}\sqrt{N_1 N_2}}{\sqrt{\pi}\sigma_0^3}\int dx (\psi_L^2-\psi_R^2)\,x\,e^{-x^2/\sigma_0^2}
\; ,\label{eta}\\
&\varepsilon=\frac{g_{12}N_1}{\sqrt{\pi}\sigma_0^3}\int_{-\infty}^{\infty}dx(\psi_L^2-\psi_R^2)\left(3\frac{x}{\sigma_0}-2\frac{x^3}{\sigma_0^3}\right)e^{-x^2/\sigma_0^2}
.\\
&\omega_\xi^2=1+\frac{g_{12}N_1}{\sqrt{\pi}\sigma_0^3}\int dx (\psi_L^2+\psi_R^2)\left(2\frac{x^2}{\sigma_0^2}-1\right)e^{-x^2/\sigma_0^2}
\; ,\label{m_xi}\\
&\omega_\sigma^2=1+\frac{3}{\sigma_0}+\frac{g_{22}N_2}{\sqrt{2\pi}\sigma_0^3}\nonumber\\
&+\frac{g_{12}N_1}{\sqrt{\pi}\sigma_0^3}\int dx (\psi_L^2+\psi_R^2)\left(2-10\frac{x^2}{\sigma_0^2}+4\frac{x^4}{\sigma_0^4}\right)e^{-x^2/\sigma_0^2}\, .
\end{align}
The expressions of $k$ (\ref{k})  and $\kappa$  (\ref{kapp}) are determined by the wave function overlap between
the left and right condensates, and thus $k \sim \kappa \ll 1$ by considering the weak link between bilateral condensates that
depends not only on atomic scattering lengths but also on the number of the condensates
with $N_2 > N_1$. Nevertheless, it will be seen that since the frequency of collectively atomic oscillations $\omega_{J\xi}
\propto \omega_x \sqrt{k}$ with $\omega_x=2\pi\times 12\text{Hz}$ we adopt, the corresponding oscillation timescale $ t \propto
1/ (\omega_x \sqrt{k})$ is constrained to be not larger than the typical lifetime of this type of the condensates of order $\sim
10 \text{s}$ \cite{Becker2008}, giving $k > 10^{-4}$ where $N_2$ cannot be too large.
The variation of the Gaussian width induces a time-dependent modification to the parameter $k+\kappa \sigma/\sigma_0$ in (\ref{ap2_z}) and (\ref{ap2_phi}). In the single BEC experiment~\cite{Smerzi1997,Raghavan1999}, the similar modification to $k$ can be achieved by the time-dependent laser intensity or the magnetic field with the possibility to  induce the Shapiro effect \cite{Grond2011}, {analog Kaptiza pendulum \cite{Boukobza2010} and even chaotic motions \cite{Abdullaev2000,Lee2001,Jiang2014,Tomkovic2017}}.
Here, since $\sigma \ll \sigma_0$, the time-dependent modification is relatively small compared with $k$ to be ignored in this work.
 Also, for simplicity, we consider that the spatial part of the wave function $\psi_1(x,t)$ is symmetric with respect to $x=0$ shown in Fig.~\ref{fig:wavefun}, thus giving $\Delta E=0$ in (\ref{ap2_phi}).
 The coupling of the population imbalance $z$ to the relative phase $\phi$  in (\ref{ap2_phi}) is given by the strength $\Lambda$ in (\ref{lambda}), a tunable value by changing the scattering lengths of atoms in  bilateral condensates.

It is known that the dynamical equations of pair variables $(z,\phi)$ in the single BEC system show an analogy to that of a nonrigid pendulum model, whose length depends on the angular momentum $z$ \cite{Raghavan1999}.
Nevertheless, in our binary BECs system, the pair variables $(z,\phi)$ are also coupled to the displacement of the wave function $\psi_2$
with the coupling constant $\eta$ in (\ref{eta}) depending on the interspecies coupling constant $g_{12}$ of  atoms, and also the wave function overlap between the bilateral and central condensates.
With the parameters shown in Fig.~\ref{fig:phasediagram0} and the chosen scattering lengths obeying $\Delta < 0$, which lead to the spatial parts of the wave functions schematically shown  in Fig.~\ref{fig:phasediagram0}, $\eta$ is of order one $\eta \sim 1.0$, and  $\varepsilon \sim 0.1$ due to cancellation between the contributions from the left and right wave functions.
In the case of the initial deviation of the wave-function width $\sigma(0)=\dot{\sigma}(0)=0$,
the time-dependent $\sigma$ is driven by $\xi$, giving $\sigma \sim \varepsilon \xi$, which in turn leads to the term $\varepsilon \sigma$ in the equation of $\xi$ (\ref{ap2_xi}) of the order of $\varepsilon^2 \xi$ with $\varepsilon^2 \ll \eta$ to be safely ignored.
%
From the viewpoint of the  wave function $\psi_2$, the presence of $\psi_R$ and $\psi_L$  will make trap potential narrower and affects slightly the natural vibration frequency in $\xi$ and give a modified frequency $\omega_\xi$ in (\ref{m_xi}).
Since we just focus on various types of coherent oscillations between bilateral condensates with the dynamical variables $z, \,\phi$, apart from their mutual coupling, they are also coupled to the displacement of the central condensate $\xi$.
 Ignoring the $\varepsilon\sigma$ term in (\ref{ap2_xi}), the small deviation from the wave-function width $\sigma$ is decoupled.
In fact, in this approximation, once the time-dependence of $\xi$ is found, the equation of $\sigma$ in (\ref{ap2_sigma}) can be solved, and then  plugging all solutions to the equations (\ref{beta}) in Appendix can find  $\alpha$ and $\beta$. Also notice that taking into account the time-dependent $\sigma$ by involving (\ref{ap2_sigma})  certainly can
improve the agreement with the full numerical results.  
Thus, after the further replacement $\xi\rightarrow\sqrt{2N_2/N_1}\xi$, the set of the key equations  can be simplified as
\begin{align}
&\dot z+2k\sqrt{1-z^2}\sin\phi=0\label{ap3_z} \, ,\\[7pt]
&\dot \phi-\Lambda z-2 k \frac{z}{\sqrt{1-z^2}}\cos\phi+ \eta \xi=0 \, ,\label{ap3_phi}\\[7pt]
&\ddot{\xi}+\omega_\xi^2\xi-\eta z=0 \, . \label{ap3_xi}
\end{align}
These are the main results of this section and we will term them the coupled pendulums (CP) dynamics.  The corresponding Hamiltonian then reads as
\begin{align}
H=\frac{\Lambda}{2}z^2-2k\sqrt{1-z^2}\cos{\phi}+\frac{1}{2}\dot{\xi}^2+\frac{\omega_\xi^2}{2}\xi^2-\eta \,z\,\xi \, .
\label{Hz3}
\end{align}
%
In our binary BECs case, the above Hamiltonian also shows an analogy between the dynamics of  coupled pendulums and
the BEC dynamics as in Ref.~ \cite{Raghavan1999}.

\section{Equilibrium Solutions and Stability analysis}\label{sec3}
%
In the two-component BEC system, the relevant equations of motion for describing quantum coherent oscillations between bilateral condensates in the presence of the central condensate are effectively described by the dynamics of two-coupled pendulums. In this case, the stationary states with the vanishing time-derivative terms now obey
\begin{align}
&2k\sqrt{1-z_0^2}\sin{\phi_0}=0,\label{z0}\\
&\Lambda z_0+2k\frac{z_0}{\sqrt{1-z_0^2}}\cos{\phi_0}-\eta\,\xi_0=0,\label{phi0}\\
&\omega_\xi^2\xi_0-\eta\, z_0=0.\label{xi0}
\end{align}
As a result, the stationary relative phase is given by
\begin{equation}
\phi_0=0\, ,\qquad\text{or}\qquad  \phi_0=\pm\pi.
\label{sol_phi0}
\end{equation}
Also, Eq. (\ref{xi0}) leads to a relation between the population imbalance $z_0$ and the displacement of the central condensate $\xi_0$
\begin{equation}
\xi_0=\omega_\xi^{-2}\eta z_0.\label{sol_xi0}
\end{equation}
Inserting the relation (\ref{sol_xi0})  into Eq. (\ref{phi0}) together with $\phi_0=0$ or $\pm \pi$,  the  population imbalance is found to be
\begin{align}
z_0=0,\qquad \text{or}\qquad z_0=\pm \sqrt{1-\frac{4k^2 }{\Lambda_{\text{eff}}^2}},
\label{sol_z0}
\end{align}
where $\Lambda_{\text{eff}}= \Lambda-\omega_\xi^{-2}\eta^2$. The solution $z_0\neq 0$ exists as long as $\Lambda_{\text{eff}}/ 2k \ge 1$ or $\Lambda_{\text{eff}}/ 2k \le -1$ in the MQST state for the  atom system. Notice that $\Lambda_\text{eff}$ can be positive or negative, and it is tunable by changing either $g_{11}$ or $g_{12}$ to vary $k$ (\ref{k}), $\Lambda$ (\ref{lambda}), and $\eta$ (\ref{eta}) .
There are four different types  of stationary states,
 $(z_0,\,\phi_0)=$ $(0,\,0)-$I, $(0,\,\pm \pi)-$II, $(\pm\sqrt{1-4k^2/\Lambda_\text{eff}^2},\,\,0 )-$III and $(\pm\sqrt{1-4k^2/\Lambda_\text{eff}^2},\,\,\pm \pi)-$IV, summarized in Table \ref{tb:config}.
Furthermore, the corresponding energy for each stationary states can be computed  through (\ref{Hz3}) using the relation of stationary-state solutions (\ref{sol_xi0}), where the results are listed in Table \ref{tb:limit}. In the case $\Lambda_{\text{eff}}/2k >1$, the ground state   is in configuration I where the bilateral condensates have the same populations $z_0=0$, thus the central condensate is centered at $\xi_0=0$, consistent with Fig.~\ref{fig:phasediagram0}. As for $\Lambda_{\text{eff}}/2k <-1$, the ground state shows the population imbalance for $z_0\neq0$ and the wavefunction $\psi_2$ is centered at $\xi_0\neq0$ due to (\ref{sol_xi0}) for a given $z_0$ also seen in Fig.~\ref{fig:phasediagram0}.

We next study the stability of the system around the equilibrium solutions. To do so, we consider the small perturbations around $\xi_0$, $z_0$, and $\phi_0$ defined by
\begin{align}
\xi(t)&=\xi_0+ \delta\xi(t), \label{deltaxi} \\
z(t)&=z_0+\delta z(t), \label{deltaz} \\
\phi(t)&=\phi_0+\delta\phi(t)\label{deltaphi}.
\end{align}
Substituting (\ref{deltaxi})--(\ref{deltaphi}) into Eqs. (\ref{ap3_z})--(\ref{ap3_xi}), and using $\xi_0=\omega_\xi^{-2}\eta z_0$, one obtains the linearized equations of motion for  $\delta\xi(t)$ and $\delta z(t)$
as
\begin{widetext}
\begin{align}
\begin{pmatrix}
\delta \ddot{\xi}(t) \\[8pt]
\delta \ddot{z}(t)
\end{pmatrix}
=
\begin{pmatrix}
-\omega_\xi^2 \quad& \eta \\[8pt]
{
\displaystyle
{2k\eta\sqrt{1-z_0^2}\cos\phi_0}}\quad
&
{\displaystyle
2k\cos\phi_0\left[\frac{(-\Lambda+2\Lambda z_0^2-\omega_{\xi}^{-2}\eta^2z_0^2)}{\sqrt{1-z_0^2}}-2k\cos{\phi_0}\right]}
\end{pmatrix}\begin{pmatrix}
\delta \xi(t) \\[8pt]
\delta z(t)
\end{pmatrix}.
\label{matrix}
\end{align}
%

\begin{table}[h]
\caption{Stationary states}
 \begin{tabular}{c c c c c}
 \hline
 \hline
 Label &  $z_0$ & $\phi_0$ &\quad $\omega_{\pm}$  \\
 \hline
 I &  0 &\quad 0 &\quad$\Big[
\omega_\xi^2+\omega_{J\xi}^2\pm \sqrt{(\omega_\xi^2-\omega_{J \xi}^2)^2+{8k\eta^2}}
\,\Big]/2$ \\[13pt]

 II & 0 &\quad $\pm\pi$ &\quad $\Big[
\omega_\xi^2+\omega_{J\xi}^2\pm \sqrt{(\omega_\xi^2-\omega_{J \xi}^2)^2-{8k\eta^2}}
\,\Big]/2$\\[13pt]

 III & $\pm\sqrt{1-4k^2/\Lambda_\text{eff}^2}$ &\quad 0 &\quad$\Big[
\omega_\xi^2+\omega_{J\xi}^2\pm \sqrt{(\omega_\xi^2-\omega_{J \xi}^2)^2+{8k\eta^2}\sqrt{1-z_0^2}}
\,\Big]/2$\\[13pt]

  IV & $\pm\sqrt{1-4k^2/\Lambda_\text{eff}^2}$ &\quad $\pm\pi$ & \quad $\Big[
\omega_\xi^2+\omega_{J\xi}^2\pm \sqrt{(\omega_\xi^2-\omega_{J \xi}^2)^2-{8k\eta^2}\sqrt{1-z_0^2}}
\,\Big]/2$ \\[6pt]
 \hline
 \hline
\end{tabular}
\label{tb:config}
\end{table}
\end{widetext}

\begin{table}[h]
\caption{Mean field extreme for different $\Lambda_\text{eff}/2k$ parameters}
 \begin{tabular}{c| c c c c }
 \hline
 \hline
 &  stable &  unstable & energy  & \\ [0.5ex]
 \hline
 $\Lambda_\text{eff}/2k>1$ & I, IV & II & $E_{\text{IV}}>E_\text{I}$ &    \\[6pt]

 $-1<\Lambda_\text{eff}/2k<1$ & I, II & none & $E_{\text{II}}>E_\text{I}$ &   \\[6pt]

 $\Lambda_\text{eff}/2k<-1$ & II, III& I & $E_{\text{II}}>E_\text{III}$ &   \\[6pt]
 \hline
 \hline
\end{tabular}
\label{tb:limit}
\end{table}
Notice that the linear term in $\delta \phi$ vanishes where $\phi_0=0$ or $\pm \pi$ for the stationary-state configurations is evaluated in obtaining the linearized equations of motion.
 Considering the oscillation motion of $\delta\xi(t)$ and $\delta z(t)$ with $\delta\xi(t),\delta z(t) \propto \exp(i\omega t)$, we obtain the eigenfrequencies
\begin{align}
&\omega_{\pm}^2=\nonumber\\
&\frac{1}{2}\left[
\omega_\xi^2+\omega_{J\xi}^2\pm \sqrt{\left(\omega_\xi^2-\omega_{J \xi}^2\right)^2+{8k\eta^2}\,\sqrt{1-z_0^2}\cos{\phi_0}}\,
\right].
\label{eigenvalue}
\end{align}

In addition to the nature frequency $\omega_\xi$ for the central condensate, there exists an
oscillating frequency
\begin{align}
\omega_{J\xi}^2
=\omega_J^2+
\frac{2k\cos\phi_0 \, \omega_\xi^{-2}\eta^2z_0^2 }{\sqrt{1-z_0^2}}\,  \label{omegaj}
\end{align}
with
\begin{equation}
\omega_{J}^2 =2k\cos\phi_0\, \left[\, 2k\cos{\phi_0}+\frac{\Lambda(1-2z_0^2)}{\sqrt{1-z_0^2}}\, \right]
\end{equation}
which manifests the nature frequency for quantum oscillations between two bilateral condensates $\omega_J$ with the effects from  the dynamics of the central condensate \cite{Smerzi1997}.
However, the true eigenfrequencies are the mixture of these two fundamental frequencies $\omega_\xi,\, \omega_{J\xi}$ as they couple. With the parameters  in the immiscible regime, typically for small $k$ where the wave-function overlap between bilateral condensates is small, this  gives $\omega_{\xi} \gg \omega_{J \xi}$.
In the two-component BECs system, the existing oscillation frequency $\omega_{\xi}$ of the central condensate will drive the dynamics of the pair variables $(z,\phi)$ into the oscillatory motions with frequencies $\omega_-$ and $\omega_+$ where $\omega_+ \gg \omega_-$.

Later we will manipulate the initial conditions of $z (0), \,\phi (0)$ as well as $\xi(0),\, \dot\xi (0)$ to effectively switch off the fast varying mode of frequency $\omega_+$, giving the results presumably  to those in Ref.~\cite{Smerzi1997}.
Then, the presence of  the fast varying mode for more general initial conditions will also be studied to see its effects on quantum coherent phenomena $(z, \phi)$. We will learn that,
 in the case of relatively small $k\sim 10^{-3}$ with very small wavefunction overlap, the time average over the evolution of the fast varying mode is adopted to find the deviation from the results in Ref.~\cite{Smerzi1997}.
Those trajectories are regular motion moving around the stable states of the system listed in Table \ref{tb:limit}. However, we also probe the parameter regime with relatively large wave-function overlap with relatively large value $k \sim 10^{-2}$ for amplifying the influence of the fast varying mode
 that can even  drive the dynamics of $z$ and $\phi$  to the chaotic behavior, if the system starts from the  unstable state also seen in  Table \ref{tb:limit}.
%
In our system, the dynamics of quantum coherence oscillations  $(z,\phi)$ is very sensitive to the tunneling energy $k$. For even larger $k$, the proposed wave-function ansatz fails to describe the evolution of $(z,\phi)$ where the dynamics of the system can only be explored by directly solving time-dependent GP equations, which is beyond the scope of the present work.

\section{Regular motion}\label{sec4}
\subsection{General solutions for small amplitude oscillations}
After studying the stationary-state configurations and examining their dynamical stability, we step forward to discuss the time evolution of the system when $z$ and $\xi$ undergo small amplitude oscillations around the stationary states. Recalling the linearized equations of motion (\ref{matrix}), the general solutions will be the superposition of two eigenfunctions,
%

\begin{align}
\delta\xi(t)=& A_{++}e^{i\omega_{+}t}+A_{+-}e^{-i\omega_{+}t}\nonumber\\
&\;\qquad\qquad+A_{-+}e^{i\omega_{-}t}+A_{--}e^{-i\omega_{-}t},
\label{ximode}\\
\delta z(t) = &B_{++}e^{i\omega_{+}t}+B_{+-}e^{-i\omega_{+}t}\nonumber\\
&\;\qquad\qquad+B_{-+}e^{i\omega_{-}t}+B_{--}e^{-i\omega_{-}t}.
\label{zmode}
\end{align}

Substituting Eqs. (\ref{ximode}) and (\ref{zmode}) into (\ref{matrix}), one gets the relations among those coefficients
\begin{align}
&A_{+,\pm}=P_{+}\,B_{+,\pm}\;, \\
&A_{-,\pm}=P_{-}\,B_{-,\pm}\;,
\end{align}
where $P_\pm$ depends on the stationary-state  solutions obtained as
\begin{align}
P_{\pm}=&\frac{-1}{{4k\eta\sqrt{1-z_0^2}\cos{\phi_0}}}\Bigg[{\omega_\xi^2-\omega_{J \xi}^2}\nonumber\\
&\qquad\pm \sqrt{\left(\omega_\xi^2-\omega_{J \xi}^2\right)^2+8k\eta^2\sqrt{1-z_0^2}\cos{\phi_0}}\,\Bigg].
\label{P}
\end{align}
According to the experimental feasibility  where an application of a magnetic gradient trap may cause the displacements of condensates \cite{McCarron2011,Eto2015,Eto2016}, we then consider the initial conditions such as
$\delta\phi(0)=0$,
and $\delta\dot{\xi}(0)=0$.
The corresponding solutions are
\begin{align}
\delta \xi(t)=&\left(\frac{-P_+P_- }{P_+-P_-}\delta z (0)+\frac{ P_+}{P_+-P_-}\delta\xi (0) \right)\cos{\omega_+t}\nonumber\\
&+\left(\frac{P_+P_-}{P_+-P_-}\delta z (0) -\frac{P_-}{P_+-P_-} \delta\xi (0) \right)\cos{\omega_-t},
\label{sol2_xi}
\end{align}
\begin{align}
\delta z(t)=&\left(\frac{-P_-}{P_+-P_-}\delta z (0) +\frac{1}{P_+-P_-}\delta\xi (0) \right)\cos{\omega_+t}\nonumber\\
&+\left(\frac{P_+}{P_+-P_-}\delta z (0)-\frac{1}{P_+-P_-}\delta\xi (0) \right)\cos{\omega_-t}.
\label{sol2_z}
\end{align}
In the case that $\psi_L$ and $\psi_R$ are very spatially separated,
resulting in small tunneling energy $k$ in  (\ref{k}), two  frequencies will have very different values, $\omega_\xi \gg\omega_{J\xi}$, since $\omega_{J\xi} \propto \sqrt{k} $ in (\ref{omegaj}).
Then the eigenfrequencies  (\ref{eigenvalue}) and the coefficients (\ref{P}) can be further approximated respectively as
\begin{align}
\omega_+\rightarrow \omega_\xi
\quad\text{,}\quad
\omega_-\rightarrow \omega_{J\xi}
\end{align}
and
\begin{align}
P_+\rightarrow -\frac{\omega_\xi^2}{2k\eta\sqrt{1-z_0^2}\cos{\phi_0}}+\mathcal{O}\left(k\right)\text{,}\,\,\,
P_-\rightarrow \frac{\eta}{\omega_\xi^2}+\mathcal{O}\left(k \right).
\label{coe}
\end{align}
The general solutions for small $k$ now become
\begin{align}
&\delta\xi(t)\simeq\nonumber\\
& \frac{1}{\omega_\xi^4}
\bigg[-\omega_\xi^2\left(\eta\,\delta z (0)-\omega_\xi^2\delta \xi (0) \right)\cos{\omega_\xi t}\nonumber\\
&+\eta\left(\omega_\xi^2\delta z (0) +2k\eta\sqrt{1-z_0^2}\cos{\phi_0}\delta\xi (0) \right)\cos{\omega_{J \xi}t}\bigg], \label{linsolxi} \\
&\delta z(t)\simeq \nonumber\\
&\frac{1}{\omega_\xi^4}\left[2k\eta\sqrt{1-z_0^2}\cos{\phi_0}\left(\eta\,\delta z(0) -\omega_\xi^2\delta \xi (0) \right)\cos{\omega_\xi t} \right. \nonumber\\
& \left.\
+\omega_\xi^2\left(\omega_\xi^2\delta z(0)+2k\eta\sqrt{1-z_0^2}\cos{\phi_0}\delta \xi (0) \right)\cos{\omega_{J\xi}t}\right].
\label{linsolz}
\end{align}

In this binary BEC system, it is important to explore the effects from the additional mode with frequency $\omega_{\xi}$ on the coherent oscillations  between bilateral condensates.
To do so, we first  choose the initial displacement of the central condensate and the initial population imbalance as ${-P_-}\delta z (0) + \delta\xi (0)=0 $ in
(\ref{sol2_xi}) and (\ref{sol2_z})  where  this choice of the initial conditions  leads to
the  amplitude  of the rapidly oscillatory motion vanishing, thus leaving with the slowly varying motion only.
For the situations of small $k$, using (\ref{coe}) we obtain a linear relation,
\begin{equation}
\delta\xi (0) = \omega_\xi^{-2}\eta\,\delta z (0)\, ,\label{dxi_dz_in}
\end{equation}
 as in the stationary-state solutions (\ref{sol_xi0}).
%
In this case, $ \delta\xi (t)$ and $ \delta z (t)$ undergo slow oscillations and the linearized solutions (\ref{linsolxi}) and (\ref{linsolz}) of the system read as
%
\begin{align}
&\xi_s (t)
\simeq \xi_0+ \omega_\xi^{-2}\eta\delta z (0) \cos\omega_{J\xi} t\; ,\label{lsol1}\\
& z_s(t)
\simeq z_0+ \delta z(0) \cos\omega_{J\xi} t. \label{lsol2}
\end{align}
Notice that together with (\ref{sol_xi0}) and (\ref{dxi_dz_in}),
%
$\xi (t) =\omega_{\xi}^{-2}\eta  z (t)  $
%
and Eqs.(\ref{ap3_z})--(\ref{ap3_xi}) reduce to
the same set of the equations in the single BEC case in Ref.~\cite{Smerzi1997} by letting
\begin{equation} \label{Lambda}
 \Lambda_\text{eff}=\Lambda-\eta^2\omega_{\xi}^{-2} \, .
\end{equation}
In addition, the solution (\ref{lsol1}) shows $\dot \xi_s  \ll \xi_s$.

Another interesting initial condition is $\delta\xi (0)=0$ and in this case we can activate properly the rapidly oscillatory motion with the solutions
\begin{align}
&\xi (t) \simeq \xi_s(t) + C_{\xi} \cos \omega_{\xi}t \;,\label{lso12a}\\
&z (t) \simeq z_s(t)  + C_{z} \cos \omega_\xi t \;,\label{lso12b}
\end{align}
where
\begin{align}
& C_\xi=-{\eta\omega_\xi^{-2}\,\delta z(0)}\;, \label{Cxi} \\
& C_z={2 \omega_{\xi}^{-4} k\eta^2\sqrt{1-z_0^2}\cos{\phi_0}\delta z(0) } \label{C} \,
\end{align}
and $C_z \ll C_\xi$ in the regime of small $k$. Later we will explore the effects from the rapidly oscillatory mode on the dynamics of $(z,\phi)$,  in the system. However, this system involves two largely separate frequencies, namely $\omega_{\xi} \gg \omega_{J\xi}$, for small $k$ with small wave-function overlap between bilateral condensates.
Although numerical trajectories  of the CP dynamics is straightforward, we find more convenient to represent the  trajectories in Poincar\'e maps (stroboscopic plots at every period, $2\pi/\omega_{\xi}$). These trajectories  in Poincar\'e maps can be analytically understood by considering  the time-averaged effects over the evolution of the  $\omega_{\xi}$ mode in the timescale $T$ where $1/\omega_{\xi} \ll T \ll 1/\omega_{J\xi}$, with which we  construct the corresponding effective potential in next section.


\subsection{The effective potential}
In order to construct an effective potential for the $(z,\phi)$ dynamics, we substitute (\ref{ap3_z}) into (\ref{Hz3}) to obtain
\begin{align}
&4k^2(1-z^2)-\dot{z}^2\nonumber\\
&\qquad=\left(\frac{\Lambda}{2}z^2+\frac{\dot\xi^2}{2}+\frac{\omega_\xi^2\xi^2}{2}-\eta\xi\,z-H(0) \right)^2,
\label{eff1}
\end{align}
%
where $H (0) $ is a constant value determined by the initial state $(z(0),\,\phi(0),\,\xi(0),\,\dot{\xi}(0))$.
To get an effective potential for the small amplitude oscillations around $z_0$ and $\xi_0$ that  includes the effects from the $\omega_{\xi}$ mode of the general $C^2$, $C^3$, and $C^4$ terms, we substitute (\ref{lso12a}) and (\ref{lso12b})
into (\ref{eff1}) giving
\begin{align}
&4k^2\left[1-(z_s+C_z\cos{\omega_\xi t})^2\right]-{(\dot{{z}_s}-\omega_\xi C_z\sin{\omega_\xi t})}^2=\nonumber\\
&\bigg[\frac{\Lambda}{2}({z_s}+C_z\cos{\omega_\xi t})^2+\frac{1}{2}(\omega_\xi C_\xi\sin{\omega_\xi t})^2+\frac{\omega_\xi^2}{2}(\eta \omega_\xi^{-2}{z_s}\nonumber\\
&+C_\xi\cos{\omega_\xi t})^2-\eta(\eta \omega_\xi^{-2}z_s+C_\xi\cos{\omega_\xi t})\,(z_s+C_z\cos{\omega_\xi t})\nonumber\\
&-H (0) \bigg]^2 \, ,
\label{veff2}
\end{align}
where, since $\dot \xi^2_s \ll \xi_s^2$, the term $\dot \xi^2_s$ is ignored for keeping the leading-order effects. Although $C_z \ll C_{\xi}$ in the small $k$ approximation, we still retain the terms of $C_z$, which are small, for seeing how they contribute to the effective potential.
Let us now introduce the time-averaged  population imbalance $z$ over the timescale $T$ for $1/\omega_{\xi} \ll T \ll 1/\omega_{J\xi}$ as
\begin{align}
\langle z_s (t)\rangle \equiv \frac{1}{T}\int_0^T dt z_s (t) = \bar z (t),
\label{aver}
\end{align}
where $\bar z (t)$ will be determined self-consistently from the resulting effective potential.
It is now straightforward to  find the expression
\begin{align}
\dot{\bar z}^2+V_\text{eff}( \bar z)=4k^2-H(0)^2\, ,
\label{dotz_eq}
\end{align}
from which the effective potential is obtained as
\begin{align}
V_\text{eff}(\bar z)=V_0(\bar{z})+\delta V(\bar{z}).\label{veff}
\end{align}
Among them, $V_0(\bar{z})$ is the well-known effective potential for a single BEC in a double-well trap potential~\cite{Smerzi1997}
\begin{align}
V_0(\bar{z})=\bar z^2 \left(4 k^2-H (0) \Lambda _{\text{eff}}+\frac{\Lambda _{\text{eff}}^2}{4} \bar z^2 \right),
\end{align}
and $\delta V(\bar{z})$ is the corrections due to the high-frequency $\omega_\xi$ mode with the terms of $C_z$ and $C_\xi$, given by
\begin{align}
&\delta V(\bar{z})\nonumber\\
&=\frac{1}{2} \left[4 k^2+\omega _{\xi }^2-\Lambda  \left(H (0)-\frac{\Lambda _{\text{eff}}}{2} \bar z^2 \right)+\Lambda _{\text{eff}}^2 \bar z^2 \right]C_z^2 \nonumber\\
&+\omega _{\xi }^2\left(\frac{\Lambda _{\text{eff}}}{2} \bar z^2  -H (0)\right)C_\xi^2+\eta\left( H (0)-\frac{\Lambda _{\text{eff}}}{2}  \bar{z}^2 \right)C_zC_\xi
\nonumber\\
&-\frac{\eta  \omega _{\xi }^2}{2} C_zC_\xi^3+\frac{1}{8} \left(2 \Lambda _{\text{eff}} \omega _{\xi }^2+5 \eta ^2\right)C_z^2C_\xi^2\nonumber\\
&-\frac{3\eta  \Lambda}{8}  C_z^3C_\xi+\frac{3 \Lambda ^2}{32}C_z^4+\frac{\omega _{\xi }^4}{4}C_\xi^4 \, .
\end{align}
Furthermore, since $\xi(t)= \omega_\xi^{-2}\eta z(t)$ holds true in the case of $C_z=C_{\xi}=0$, the whole dynamics of $(z,\phi,\xi)$ can reduce to that of $(z,\phi)$ with symmetry of $\Lambda_{\text{eff}} \rightarrow -\Lambda_{\text{eff}}$, and $\phi \rightarrow -\phi +\pi$ as in Ref.~ \cite{Smerzi1997}.
Nonetheless, since $C_z$ and $ C_{\xi}$ are in general nonzero, the system then does not obey the above-mentioned symmetry. We will verify this in the later numerical studies.

\subsection{Results and Discussions}
\subsubsection{$\Lambda_\textsf{eff}/2k>2$ (Josephson oscillation, running phase and $\pi$-mode self trapping)}
Consider the numbers of atoms in this binary condensates with the atom numbers $N_1=200,\,N_2=1000$. In Fig.~\ref{fig:phaseportrait}, we have chosen scattering lengths $a_{11}=97.5a_{0},\,a_{22}=50a_{0},\,a_{12}=85a_{0}$.
As stated previously, we prepare the initial states of the condensates, the Gaussian-like spatial functions $\psi_L (x)$ and $\psi_R (x)$  together with the initial wave function of the central condensate $\psi_2(x,t=0)$ with width $\sigma_{0}$ determined numerically from finding the stationary-state solutions of GP equations (\ref{GPe1}) and (\ref{GPe2})  with the forms shown in Fig.~\ref{fig:wavefun}.
The resulting wave functions can in turn give the values of parameters $k=0.0025\,$, $\Lambda=1.91$, and $\eta=1.25$  via (\ref{k}), (\ref{lambda}), and (\ref{eta}), which lead to $\Lambda_\text{eff}/2k=68 >2$.

We first  depict the trajectories by directly solving the equations of motions (\ref{ap3_z})--(\ref{ap3_xi}) where  the initial conditions are slight away from the stationary-state of $\phi_0=0$ and $z_0=\xi_0=0$ shown in Fig.~\ref{fig:phaseportrait}(a).
The initial conditions are chosen as $\delta \phi (0) =\delta \dot{\xi}(0)=0$ together with various choices of $\delta z(0)$ and $\delta \xi (0)$, obeying $\delta\xi (0) = \eta\omega_\xi^{-2}\delta z (0)$, with which  only the $\omega_{J \xi}$ mode becomes relevant.
%
Thus, by varying $\delta z(0)$ and $\delta \xi (0)$ accordingly, the plot shows the regimes of Josephon-like oscillations for $\bar z (0)=z_0 +\delta z(0) < \bar z_c (0)$, and MQST with the running phase for $\bar z(0) > \bar z_c (0) $ as in Ref.~\cite{Smerzi1997}.
The critical $\bar z$ in this case is $\bar z_c (0) = 0.24$, obtained from (\ref{zc_lambda_ge2}) below.
Also, we see the consistency of the solutions  (\ref{linsolxi}) and (\ref{linsolz}) with those of the equations in (\ref{ap3_z})--(\ref{ap3_xi}). In the case of $C_z=C_{\xi}=0$, the effective potential (\ref{veff}) reduces to that in  Ref.~\cite{Smerzi1997} drawn in Fig.~\ref{fig:phaseportrait}(c) of a double-well form.
Thus,  for $\bar z (0)  < \bar z_c (0)$ the system has large  enough kinetic energy $\dot {\bar z}^2$ to move forward and backward between two potential minima around the stationary state $z_0=0$ and $\phi_0=0$ undergoing the Josephson oscillations.
As $\bar z(0)$ increases to $\bar z(0)=\bar z_c(0)$, the initial kinetic energy of the system drives  $\bar z$  rolling down toward  one of the potential minima and then climbing up the potential hill to reach the state of $\bar z=0$ just with $\dot { \bar z}=0$, obeying the condition
\begin{align}
\dot{ \bar z}_c(0)^2+V_\text{eff}( \bar z_c (0))=V_\text{eff}(\bar z=0)\, .
\label{dotz_eq2}
\end{align}
\begin{figure}[t]
\begin{center}
\includegraphics[width=1\linewidth]{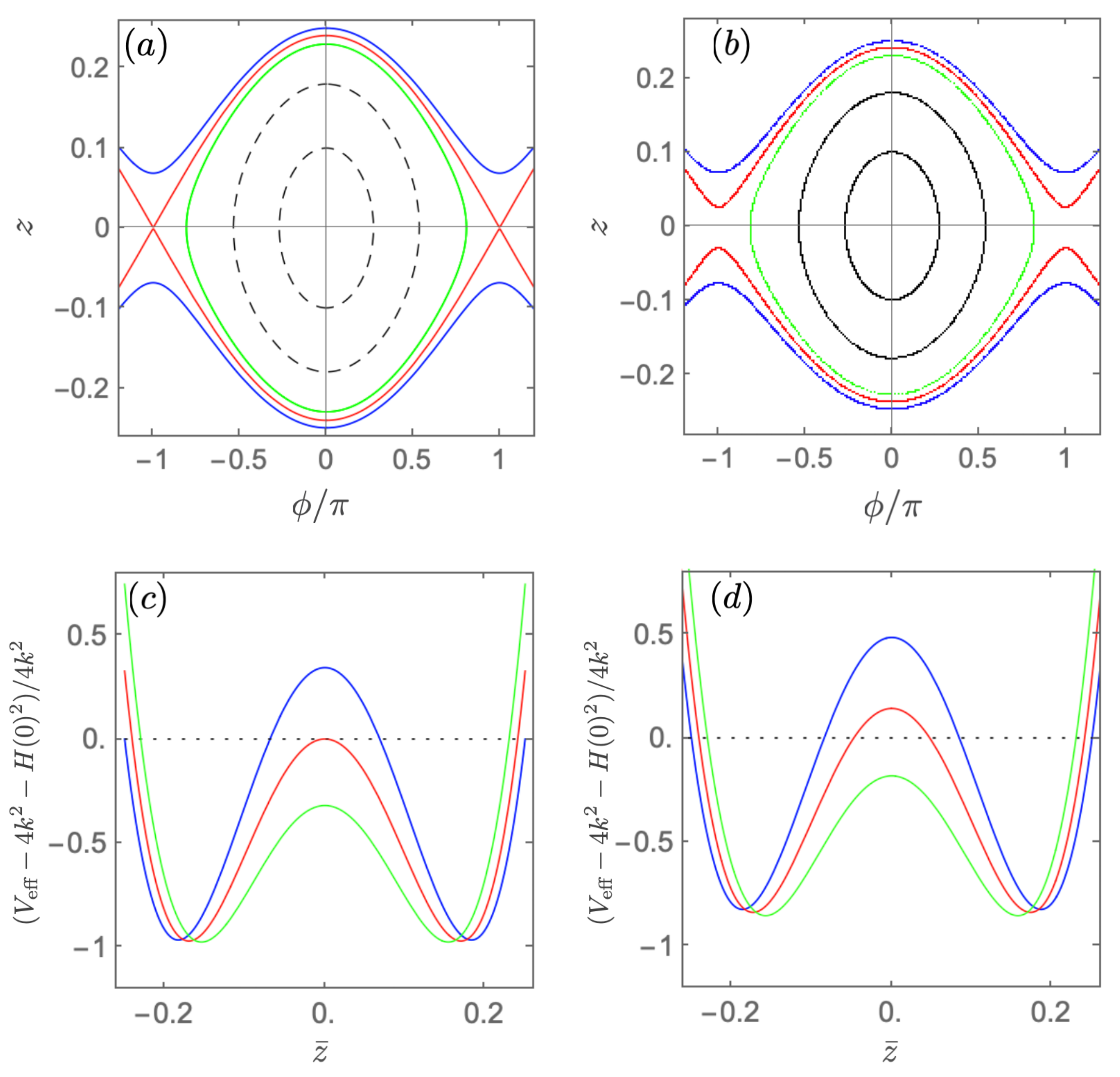}
\caption{The phase portrait of the population imbalance $z$ and the  phase difference between the bilateral condensates $\phi$ as well as the plots of the effective potential with the parameter $\Lambda=1.91,\,\eta=1.25,\,k=0.0026 $, giving $\Lambda_\text{eff}/2k=68.02)$: (a) with the initial conditions, $\delta z (0)=0.25,\,\delta \xi(0)=0.31$ [blue (gray)], $\delta z(0)=0.24,\,\delta \xi(0)=0.30$ [red (dark gray)], and  $\delta z(0)=0.23,\,\delta \xi(0)=0.29$ [green (light gray)]
where the $\omega_{\xi}$ mode is effectively switched; (b) with the initial conditions  $\delta z (0)=0.25, \delta\xi (0)=0$ [blue (gray)],  $\delta z (0)=0.24,\,\delta \xi (0)=0$ [red (dark gray)], and  $\delta z(0)=0.23,\,\delta \xi (0)=0$ [green (light gray)] where the the $\omega_{\xi}$ mode is effectively activated. Other initial conditions for both cases are $\delta\phi(0)=0$, $\dot{\xi}(0)=0$. The plots (c) and (d) are drawn for the corresponding effective potential  for (a) and  (b), respectively.}
\label{fig:phaseportrait}
\end{center}
\end{figure}
Moreover, when $\bar z(0) > \bar z_c (0)$, the relatively small kinetic energy constrains the system to move around one of the potential minima with $z_0 \neq 0$, leading to the MQST, also shown in Fig.~\ref{fig:phaseportrait}(c).

Considering the initial conditions mentioned above
together with
the population imbalance $ \bar z(0)= z_0 +\delta z (0)$ and the displacement of the central condensate $ \xi (0)=\xi_0 +\delta \xi (0)$ obeying the relation $\xi (0)=\eta \omega_{\xi}^{-2} \bar z (0)$,
$H(0)$
becomes
\begin{align}
H(0)=\frac{\Lambda_\text{eff}}{2} \bar z (0)^2-2k\sqrt{1-\bar z(0)^2}\cos{\phi(0)} \, .
\end{align}
Then, using Eq.~(\ref{ap3_z}) at the initial time to replace $\dot{ \bar z}_c(0)$ in   (\ref{dotz_eq2}), for a given $\Lambda_\text{eff}$, one can find the value of $\bar z_c(0)$  with the effective potential (\ref{eff1}) by setting $C_z=C_{\xi}=0$, which gives the known result in Ref.~\cite{Smerzi1997}.
For $\Lambda_{\text{eff}} / 2k >2$, if the initial relative phase $0\le\vert{\phi(0)}\vert\le\pi/2$, then
\begin{align}
\bar z^2_c(0)=&
\frac{4 k}{\Lambda_{\text{eff}}^2} \bigg[ \Lambda_{\text{eff}} -2 k\cos \phi(0)\nonumber\\
&\,\,+\sqrt{\Lambda_{\text{eff}} \left(\Lambda_{\text{eff}}- 4k\right) \cos^2 \phi(0)+ 4k^2 \cos^4 \phi(0) }\bigg],\nonumber\\
\label{zc_lambda_ge2}
\end{align}
and when $\pi/2\le\vert{\phi(0)}\vert\le\pi$ we have
\begin{align}
\bar z^2_c(0)=&
\frac{ 4 k}{\Lambda_{\text{eff}}^2} \bigg[ \Lambda_{\text{eff}} -2 k\cos \phi(0)\nonumber\\
&\,\,- \sqrt{\Lambda_{\text{eff}} (\Lambda_{\text{eff}}- 4k) \cos^2 \phi(0)+ 4k^2 \cos^4 \phi(0) }\bigg].
\label{zc_lambda_ge3}
\end{align}
The critical value of $\bar z_c (0)$ as a function of $\phi(0)$ draws a boundary between  the Josephson oscillations and the MQST mode to be discussed in the later numerical studies in Fig.~\ref{fig:phase_general}.
As for a negative $\Lambda_{\text{eff}}$, the solutions of $\bar z^2_c(0)$ are given due to the following symmetry $\Lambda_{\text{eff}} \rightarrow -\Lambda_{\text{eff}}$ and $\phi \rightarrow -\phi+\pi$.
The value of $\bar z_c (0)$ found in Fig.~\ref{fig:phaseportrait}(a) and \ref{fig:phaseportrait}(c) can also been achieved from (\ref{zc_lambda_ge2}) and (\ref{zc_lambda_ge3}) with $\phi(0)=0$ for the positive value of $\bar z_c(0)$.

\begin{figure}[t]
\begin{center}
\includegraphics[width=1\linewidth]{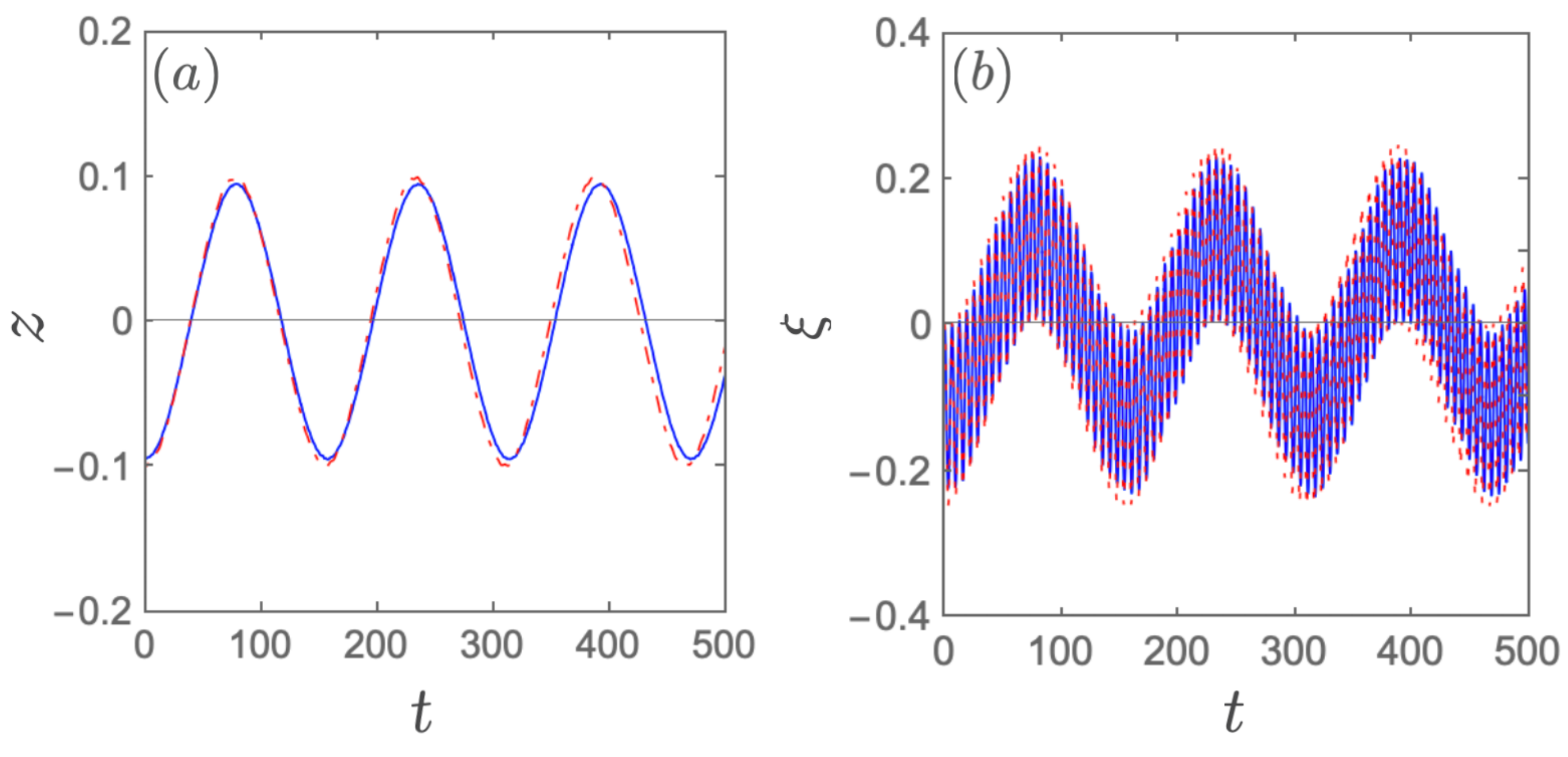}
\caption{The evolution of the population imbalance $z$ (a) and the displacement of the central condensate (b) as a function of time in the case of the Josephson oscillations are compared between the results of
the numerical solutions of the time-dependent GP equations (blue line), and the linearized equations in  Eqs.~(\ref{ap3_z})--(\ref{ap3_xi}) (red dashed line) with the same parameters described in the text and the initial conditions $z(0)=-0.1, \phi(0)=\xi(0)=\dot \xi (0)=0$.}
\label{fig:gaussian}
\end{center}
\end{figure}

In Fig.~\ref{fig:phaseportrait}(b), we present the evolutions of $ z(t)$ and $ \phi (t)$, which include the effects from the $\omega_{\xi}$ mode with the same set parameters discussed previously (see figure caption for details).
To make a comparison with the evolution in Fig.~\ref{fig:phaseportrait}(a) of the slowly varying mode, it finds more convenient to represent the  trajectories in Poincar\'e maps (stroboscopic plots at every period $T = 2\pi/\omega_{\xi}$) by collecting the results from the section of $\xi(0)=0$ and $\dot \xi (0)>0$.
It in turn can be further analyzed in terms of the effective potential as we will see later by considering the time-averaged effects from the fast oscillatory mode.
The similar Josephson-like oscillations  $( \bar z(0) < \bar z_c(0))$  and the running phase MQST dynamics $\bar z(0) > \bar z_c(0)$  are shown with a transition between them at the shifted critical value  $\bar z_{c}(0)=0.238$.
The shifted critical value $\bar z_c(0)$ can be obtained from
the conserved quantity
\begin{align}
H(0)=&\frac{\Lambda_\text{eff}}{2} \bar z (0)^2-2k\sqrt{1-(\bar z(0)		 +C_z)^2}\cos{\phi(0)}\nonumber\\
&+\frac{(\Lambda_\text{eff}+\eta\omega_\xi^{-2})}{2}C_z^2+\Lambda_\text{eff}\bar z(0) C_z +\frac{\omega_\xi^2}{2}C_\xi^2\nonumber\\
&-\eta \, C_zC_\xi,
\end{align}
where  we have substituted the results of (\ref{lsol1}) and (\ref{lsol2}) into (\ref{Hz3}), and  evaluated the Hamiltonian at an initial time.
Again, substituting  (\ref{ap3_z}) at an initial time into (\ref{dotz_eq2}) and with the help of effective potential (\ref{veff}), we can obtain the critical value of $\bar z_c (0)$ through the expression as follows
\begin{widetext}
\begin{align} \label{zc_C}
\Lambda_{\text{eff}}=&\frac{1}{\bar z_c (0)^2+4 C_z \bar z_c(0)- C_z^2}\Bigg\{4k\sqrt{1-(\bar z_c (0)		+C_z)^2}\cos{(\phi(0))}+\eta C_zC_\xi-\frac{\eta^2\omega_\xi^{-2}}{2} C_z^2 \nonumber\\&+\Bigg[\left(4k\sqrt{1-(\bar z (0)+C_z)^2}\cos{(\phi(0))}+\eta \, C_zC_\xi-\frac{\eta^2\omega_\xi^{-2}}{2} C_z^2\right)^2+\frac{16k^2}{\bar z_c (0)^2}(\bar z_c(0)^2\nonumber\\
&+4C_z \bar z_c(0)-C_z^2)[1-C_z(2\bar{z}_c(0)+C_z)-(1-(\bar z_c(0)+C_z)^2)\cos^2{(\phi (0))}]\Bigg]^{1/2}\Bigg\},
\end{align}
\end{widetext}
where  $C_z$ and $C_\xi$ are given explicitly in terms of $z_0$ and $\delta z_c (0)$ via (\ref{C}) and
$\bar z_c(0)=z_0+\delta z_c(0)$ in (\ref{lsol2}). For a fixed $\Lambda_{\text{eff}}$,  the shifted value of $\bar z_{c} (0)$  obtained from  (\ref{zc_C}) is consistent with the numerical result in Fig.~\ref{fig:phaseportrait}(b).

In Fig.~\ref{fig:gaussian} we provide numerical comparison on the results of the coupled pendulums equations (\ref{ap3_z})--(\ref{ap3_xi}) and those by solving the full time-dependent GP equations (\ref{GPe1}) and (\ref{GPe2}). The initial wave functions for both condensates, as previously said,
are prepared from the stationary solutions of GP equations with the form similar to Fig.~\ref{fig:wavefun}.  The wave  function of the central condensate can be approximated by (\ref{psi2}) with $\xi(0)=\alpha (0)=\beta (0)=0$ giving $\dot{\xi} (0)=0$.
However, for the bilateral condensates, their initial wave functions are prepared for the functions of $x$ with the relative phase $\phi(0)=0$ and then
tune the wave functions to give $z(0)\ne 0$.
In Fig.~\ref{fig:gaussian}, using the same parameters and the initial conditions we find consistency between the solutions from analogous  CP dynamics and the time-dependent GP equations.

\begin{figure}[t]
\begin{center}
\includegraphics[width=1\linewidth]{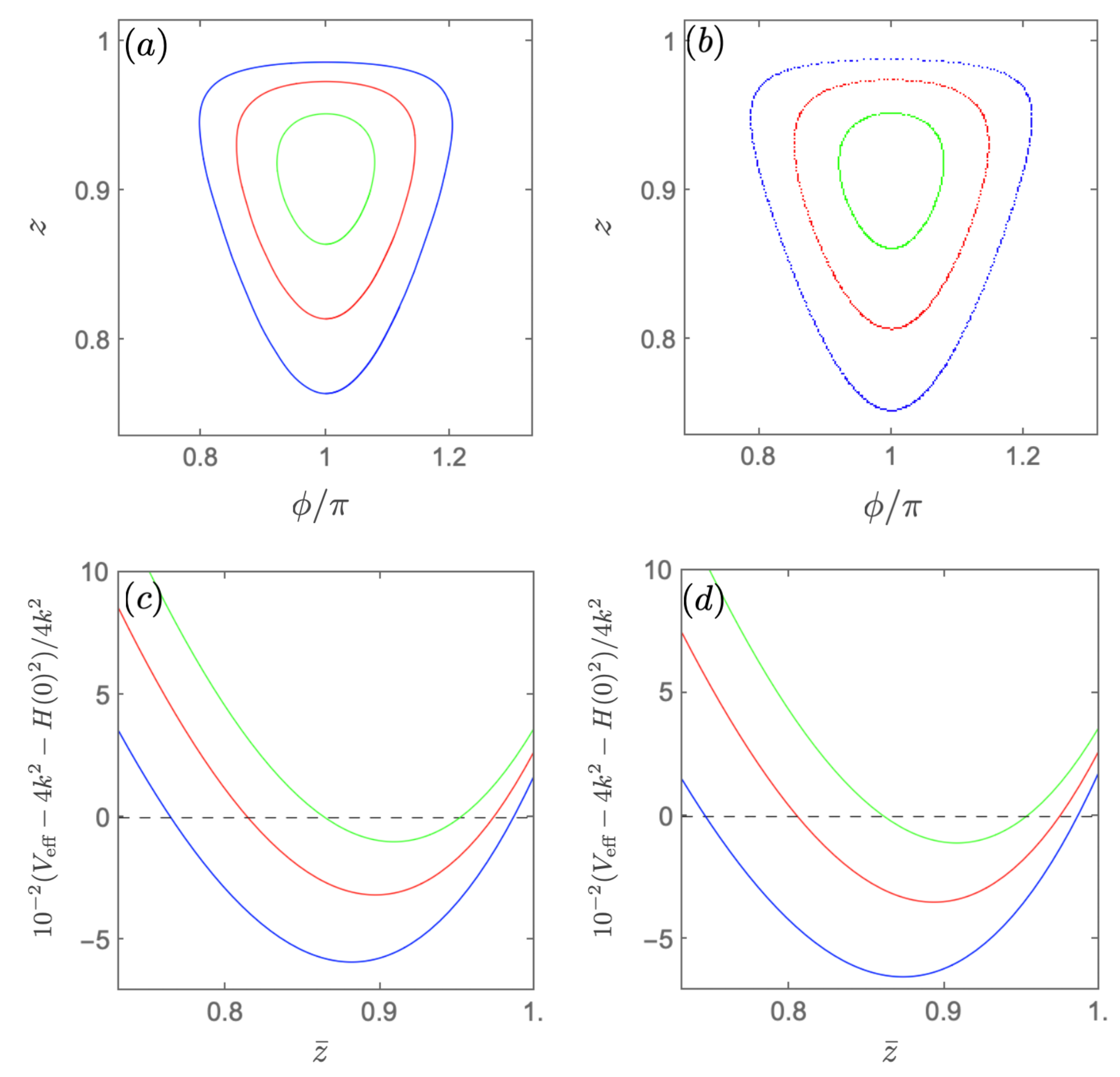}
\caption{{The phase portrait with the parameters $\Lambda=1.80$, $\eta=1.31$, $k=0.017\,(\Lambda_\text{eff}/2k=2.47)$.  The $\pi$ mode trajectories circle the stationary point $z_0=0.91,\, \phi_0=\pm\pi,\,\xi_0=1.19$ shown
in top
panels: (a) with the initial conditions, $\delta z(0)=0.15,\,\delta \xi(0)=0.20$ [blue, (gray)], $\delta z(0)=0.10,\,\delta \xi(0)=0.13$ [red (dark gray)], and  $\delta z(0)=0.05,\,\delta \xi(0)=0.065$ [green (light gray)]
where the $\omega_{\xi}$ mode is effectively switched: (b) with the initial conditions  $\delta z(0)=0.15, \delta\xi(0)=0$ [blue (gray)],  $\delta z(0)=0.10,\,\delta \xi(0)=0$ [red  (dark gray)], and  $\delta z(0)=0.05,\,\delta \xi(0)=0$ [green (light gray)] where the the $\omega_{\xi}$ mode is effectively activated. Other initial conditions for both cases are $\delta\phi(0)=0$, $\delta\dot{\xi}(0)=0$. In bottom
panels, the plots (c) and (d) are drawn for the corresponding effective potential  for (a), and  (b) respectively.}}
\label{fig:phase_pi2}
\end{center}
\end{figure}

\begin{figure}[h]
\begin{center}
\includegraphics[width=1\linewidth]{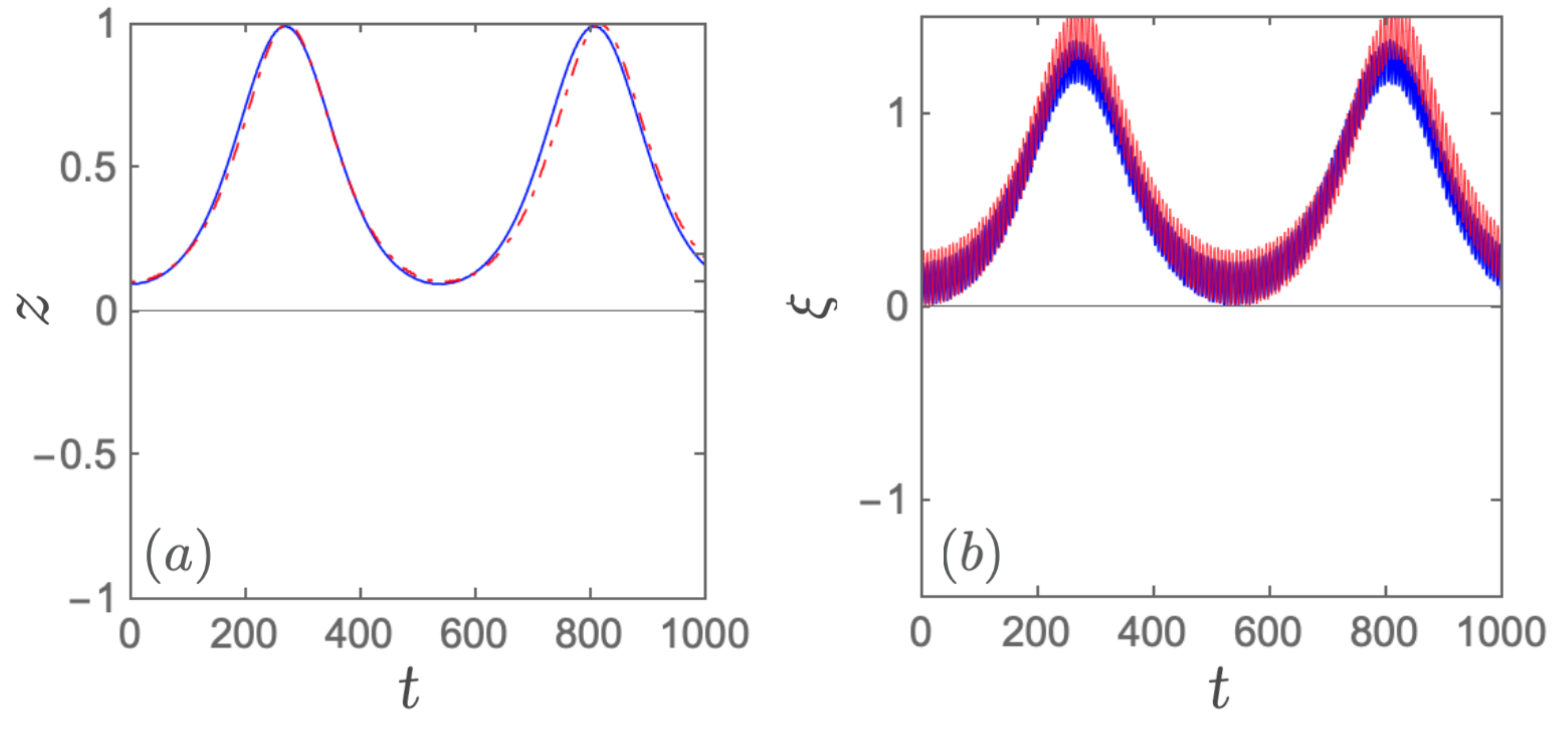}
\caption{The evolution of the population imbalance $z$ (a) and the displacement of the center condensate (b) as a function of time in the case of the MQST are compared between the results of
the numerical solutions of the time-dependent GP equations (blue line), and the linearized equations in  Eqs. (\ref{ap3_z}), (\ref{ap3_phi}), and (\ref{ap3_xi}) (red dashed line) with the parameters $a_{11}=87a_{0},\,a_{22}=50a_{0},\,a_{12}=85a_{0}$ ($\Lambda=2.0,\,k=0.007,\,\eta=1.4$) and the initial conditions are $z(0)=0.1, \,\phi(0)=\pi,\,\xi(0)=\dot \xi (0)=0$.}
\label{fig:gaussian2}
\end{center}
\end{figure}

In Fig.~\ref{fig:phase_pi2}, the scattering lengths are slightly changed to probe  the dynamics of $z,\phi$ around the stationary state $\phi_0=\pi, z_0 =0.91 \, ( \ne 0) $ obtained from (\ref{sol_z0})
with the parameters $\Lambda=1.63$, $\eta=1.267$, $k=0.005$, giving $\Lambda_\text{eff}/2k=2.5>2$.
We have the $\pi$-mode MQST oscillations in this case.
The initial conditions of $\delta z(0)$ are chosen being $\delta z(0) \ll 1$ with small perturbations around nonzero $z_0$, where the effective potential constructed in (\ref{veff}) can safely be applied.
The plots of Figs.~\ref{fig:phase_pi2}(c) and \ref{fig:phase_pi2}(d) of the respective effective potentials illustrate the single-well profile with the effects of turning on or off the $\omega_{\xi}$ mode can satisfactorily interpret the dynamics of the population imbalance $z$ shown in Fig.~\ref{fig:phase_pi2}(a) and \ref{fig:phase_pi2}(b).
This indicates that the MQST mode will persist for all range of $z(0)$ with $-1 < z (0) <1$.
The effects from nonzero $C_z$ and $C_{\xi}$ seem changing slightly the turning points of $\bar z$, as $\bar z$ oscillates around $z_0$.
For  this $\Lambda_{\text{eff}}$ with $\phi (0)=\pi$, $\bar z_c (0)=0$ is found.
This means that for  $-1< z (0) <1$, the MQST mode sets in. In general, we can find $\bar z_c(0)$ as a function of $\phi(0)$ using the effective potential obtained above to be shown in Fig.~\ref{fig:phase_pi2}. Similar check is done using the time-dependent GP equation as in Fig.~\ref{fig:gaussian} for the case of  MQST whereas
in Fig.~\ref{fig:gaussian2},  large discrepancy is found due to the fact that the two modes approach in (\ref{fig:wavefun}) may not provide a relatively good basis to parametrize the spatial parts of the wave functions of bilateral condensates.

\begin{figure}[thb]
\begin{center}
\includegraphics[width=0.9\linewidth]{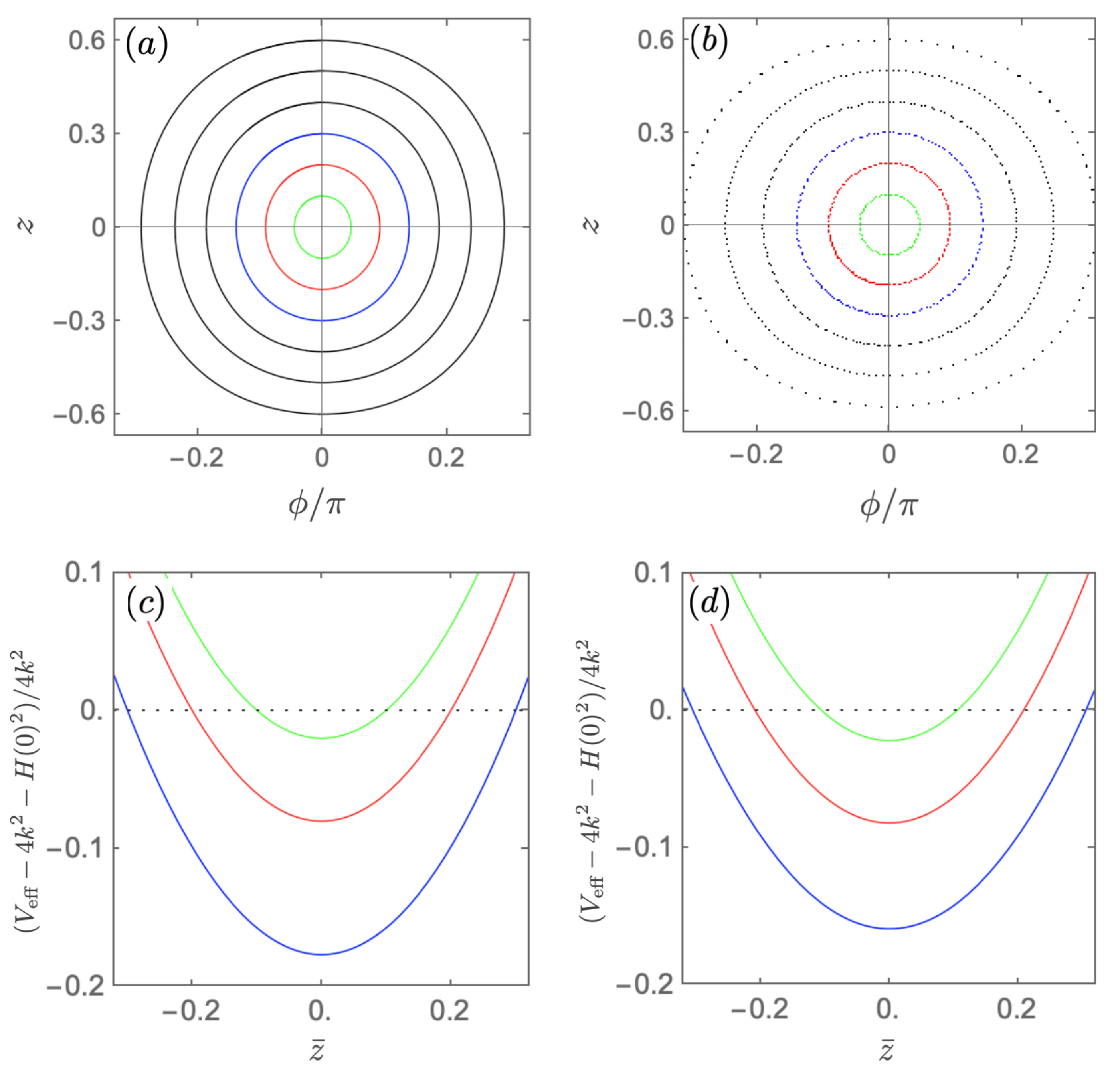}
\caption{{In top panels, the phase portrait with the parameters  of $\Lambda=1.63$, $\eta=1.27$, $\eta=0.005\,(\Lambda_\text{eff}/2k=1.07)$ is plotted for the Josephson oscillation trajectories around the stationary state: $z_0=0\, \phi_0=0,\,\xi_0=0$ (a) with the initial conditions, $\delta z (0)=0.3,\,\delta \xi (0)=0.38$ [blue (gray)], $\delta z (0)=0.2,\,\delta \xi (0)=0.25$ [red (dark gray)], and  $\delta z (0)=0.10,\,\delta \xi (0)=0.12$ [green (light gray)]
where the $\omega_{\xi}$ mode is effectively switched: (b) with the initial conditions  $\delta z (0)=0.30, \delta\xi (0)=0$ [blue (gray)],  $\delta z (0)=0.20,\,\delta \xi (0)=0$ [red (dark gray)], and  $\delta z(0)=0.10,\,\delta \xi(0)=0$ [green (light gray)] where the $\omega_{\xi}$ mode is effectively activated. Other initial conditions for both cases are $\delta\phi (0)=0$, $\delta\dot{\xi}(0)=0$. In bottom panels, the plots (c) and (d) are drawn for the corresponding effective potential  for (a), and  (b) respectively.}}
\label{fig:phase_0}
\end{center}
\end{figure}

\begin{figure}[thb]
\begin{center}
\includegraphics[width=0.9\linewidth]{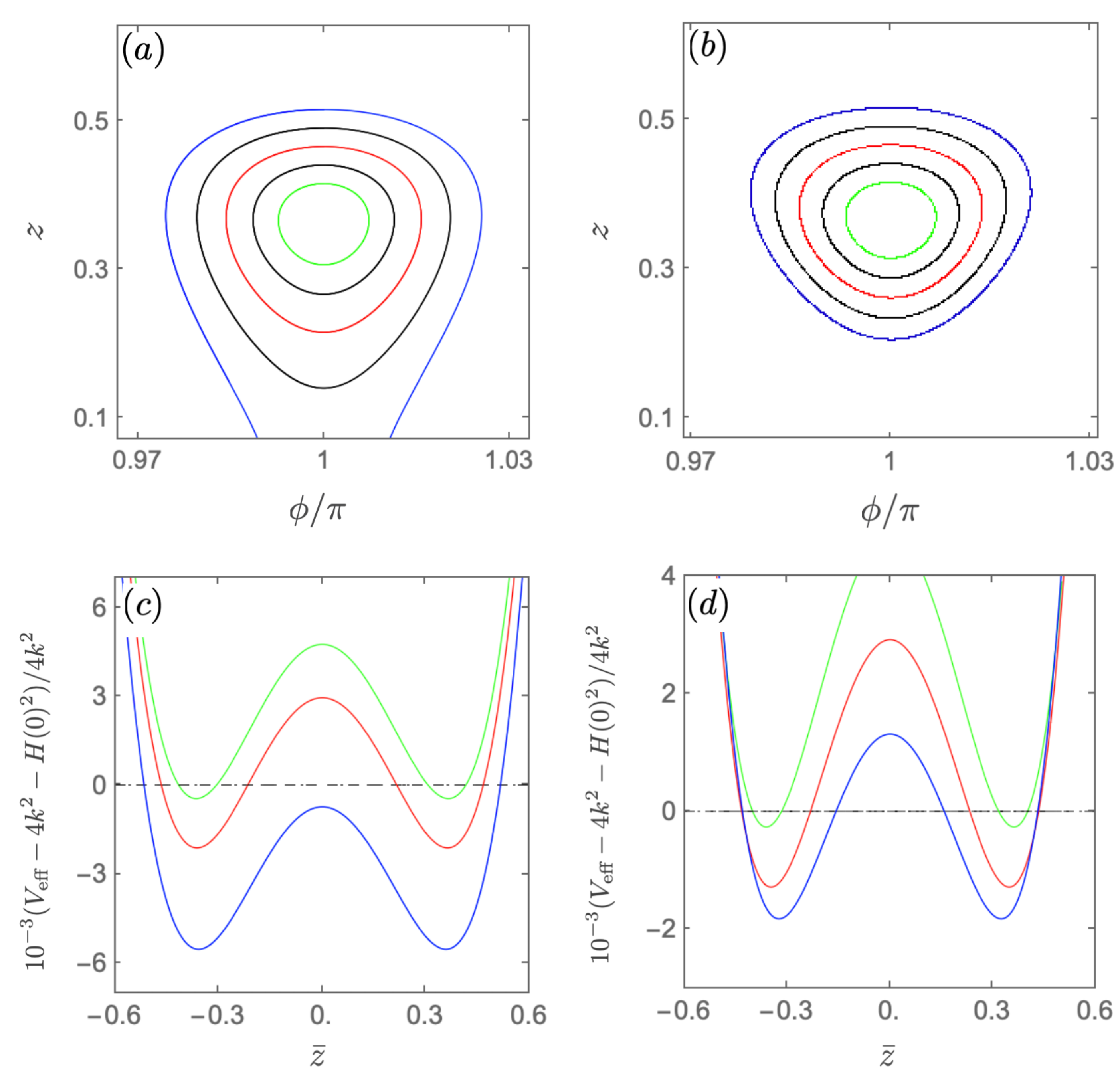}
\caption{{In top panels, the phase portrait with parameters as in Fig.\ref{fig:phase_0} is plotted for the $\pi$ mode trajectories around the stationary state: $z_0=0.37,\, \phi_0=\pi,\,\xi_0=0.47$ (a) with the initial conditions, $\delta z(0)=0.15,\,\delta \xi(0)=0.1905$ [blue (gray)], $\delta z(0)=0.100,\,\delta \xi(0)=0.127$ [red  (dark gray)], and  $\delta z(0)=0.050,\,\delta \xi(0)=0.064$ [green (light gray)]
where the $\omega_{\xi}$ mode is effectively switched: (b) with the initial conditions  $\delta z(0)=0.150, \delta\xi(0)=0$ [blue (gray)],  $\delta z(0)=0.100,\,\delta \xi(0)=0$ [red (dark gray)], and  $\delta z(0)=0.050,\,\delta \xi(0)=0$ [green (light gray)] where the the $\omega_{\xi}$ mode is effectively activated. Other initial conditions for both cases are $\delta\phi(0)=0$, $\delta\dot{\xi}(0)=0$. In bottom panels, the plots (c) and (d) are drawn for the corresponding effective potential  for (a), and  (b) respectively.}}
\label{fig:phase_pi}
\end{center}
\end{figure}

\subsubsection{$1<\Lambda_\textsf{eff}/2k<2$ (Josephson oscillation and $\pi$-mode self trapping)}
In the interval $1<\Lambda_{\text{eff}} / 2k<2$, there exist two solutions of $\bar{z}^2_c(0)$, if $\pi/2\le\vert{\phi(0)}\vert\le\pi$,
\begin{align} \label{zc_lambda_le2}
\bar z^2_c(0)=
&\frac{ 4 k}{\Lambda_{\text{eff}}^2} \bigg[ \Lambda_{\text{eff}} -2 k\cos \phi(0) \nonumber\\
&
\pm \sqrt{\Lambda_{\text{eff}} (\Lambda_{\text{eff}}- 4k) \cos^2 \phi(0)+ 4k^2 \cos^4 \phi(0) }\bigg].
\end{align}
In Figs.~\ref{fig:phase_0} and \ref{fig:phase_pi}, the parameter $\Lambda_{\text{eff}}/2k$ is tuned to the value between $1< \Lambda_{\text{eff}}/2k <
2$. On the one hand, the Josephson oscillations are seen for $\phi (0)=0$, and however, the running phase MQST does not exist in this case in Fig.~\ref{fig:phase_0} where there is no such a $\bar z_c(0)$ found  with the single-well potential, shown in (\ref{zc_C}) and (\ref{zc_lambda_le2}). On the other hand, the transition between the Josephson oscillations and the MQST  can occur in $\phi (0)= \pi$ with the double-well potential, giving the critical value of $\bar z_c (0)$ either  by (\ref{zc_lambda_le2}) for the positive $\bar z_c(0)$ with the $\omega_{J \xi}$ mode only, or by (\ref{zc_C}) involving the contributions from the $\omega_{\xi}$ mode.

\subsubsection{$0<\Lambda_\textsf{eff}/2k<1$ (Josephson oscillation)}
As for $0 < \Lambda_{\text{eff}}/2k<1 $, since there does not exist the  solution of $\bar z_c (0)$ due to the fact that the corresponding effective potential for both $\phi(0)=0$ and $\phi(0)=\pi$ shows a single well like the case in Fig.~\ref{fig:phase_0},  we find the Josephson oscillations for both relative phase difference cases.

To summarize the discussion given above, we show the phase portraits in Fig.~\ref{fig:phase_general} from  solving the double pendulum equations (\ref{ap3_z})--(\ref{ap3_xi}).
The initial conditions we choose in the plots of the top row are again $z(0)=z_0 +\delta z(0), \phi(0)=\phi_0$ and $\xi(0)=\xi_0 +\delta \xi (0),\, \dot \xi(0)=0$, where $\delta \xi(0)= \omega_{\xi}^{-2}\eta \delta z(0)$
so that the mode of $\omega_{\xi}$ is effectively switched off.
The bottom row figures  correspond to the initial condition of $\delta \xi (0)=0$ instead with the solutions that  the mode of $\omega_{\xi}$ is effectively switched on for comparison.
For top row results, in Fig.~\ref{fig:phase_general}(e), there exists a transition between the Josephson oscillations and the running phase MQST at $\phi(0)=0$ for $\Lambda_{\text{eff}}/2 k >2$, and in Fig.~\ref{fig:phase_general}(d),  the transition  occurs between the MQST and Josephson oscillations at $\phi(0)= \pm \pi$ instead apart from the Josephson oscillations at $\phi (0)=0$ for $1< \Lambda_{\text{eff}}/2k <2$, and further in Fig.~\ref{fig:phase_general}(c) the  Josephson oscillations at $\phi(0)=0, \pm \pi$ respectively  occur for $ 0<\Lambda_{\text{eff}}/2k <1$.
For $ \Lambda_{\text{eff}}/2k < 0$, with $C_{\xi}=C_z=0$, all results for $ \Lambda_{\text{eff}}/2k > 0$ are shifted from $\phi \rightarrow -\phi +\pi$.
Nevertheless, including the effects from the $\omega_{\xi}$ mode, such symmetry mentioned above does not exist.

\begin{figure*}[thb]
\begin{center}
\includegraphics[width=0.9\linewidth]{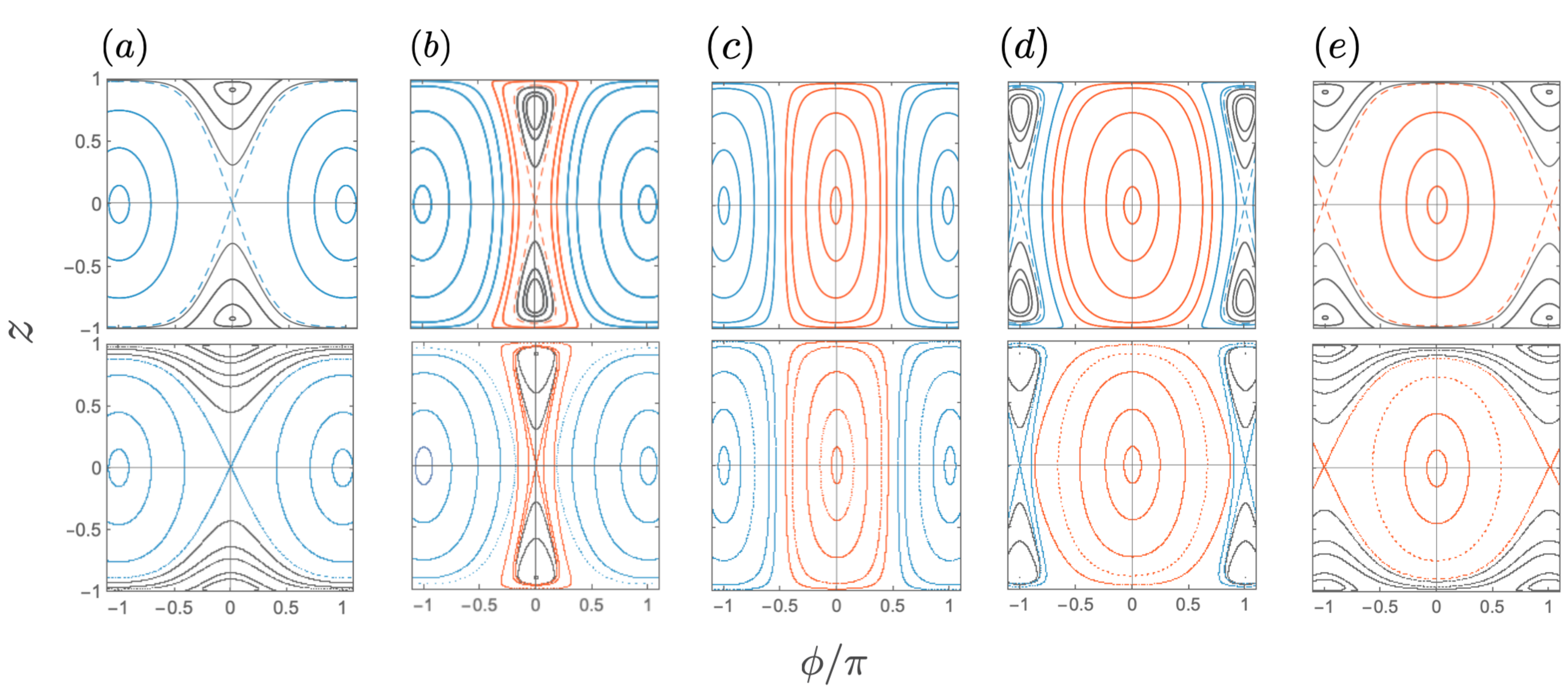}
\caption{The phase portraits, choosing different $\eta$  with fixed parameters of $\Lambda=1.35,\,k=0.005$ and $\omega_\xi=1$ by solving (\ref{ap3_z}), (\ref{ap3_phi}), and (\ref{ap3_xi}).  The top row figures correspond to the initial conditions: $z(0)=z_0 +\delta z(0), \phi(0)=\phi_0$, and $\xi(0)=\xi_0 +\delta \xi (0), \dot \xi(0)=0$, where $\delta \xi(0)=\eta \omega_{\xi}^2 \delta z(0)$ away from the various stationary states by changing $\delta z (0)$
with the solutions that the mode of $\omega_{\xi}$ is effectively switched off.
 The bottom row figures  correspond to the initial condition of $\delta \xi (0)=0$ in stead with the solutions that  the mode of $\omega_{\xi}$ is effectively switched on for comparison.
The parameters we choose are as follows
: (a) $\Lambda_\text{eff}/2k=-2.47$, $\eta=1.172$; (b) $\Lambda_\text{eff}/2k=-1.55$, $\eta=1.168$; (c) $\Lambda_\text{eff}/2k=0$, $\eta=1.162$; (d) $\Lambda_\text{eff}/2k=1.55$, $\eta=1.155$; (e) $\Lambda_\text{eff}/2k=2.47$, $\eta=1.151$. }
\label{fig:phase_general}
\end{center}
\end{figure*}


\section{Chaotic dynamics}\label{sec5}
In  previous sections, we mainly focus on various regimes of the regular motions. In particular, when two frequencies $\omega_+$ and $\omega_-$ have large difference in magnitude, the dynamics of the rapidly varying mode is averaged out giving a mean effect to  the mode of slow oscillations.
However, as the difference in magnitude between two frequencies becomes small with a relatively larger overlap between bilateral condensates, the system  exhibits chaotic phenomena with the scattering lengths in the regimes marked in Fig.~\ref{fig:phase_diagram_chaos}.
\begin{figure}[t]
\begin{center}
\includegraphics[width=1\linewidth]{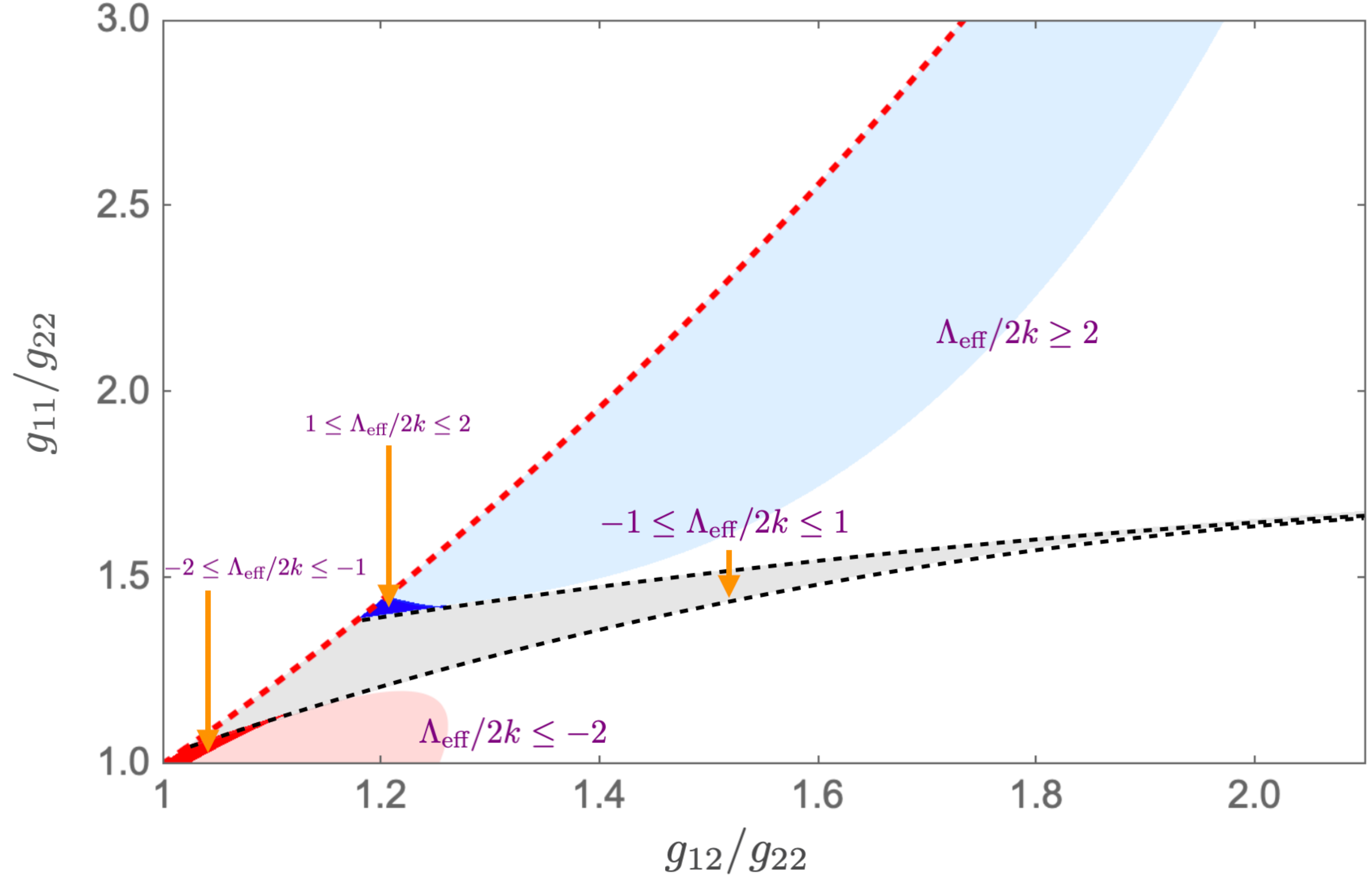}
\caption{ With same parameters of Fig.~\ref{fig:phasediagram0}, the (blue and deep blue) regimes of the scattering lengths
 lead to the chaotic behavior of the system  for  $\Lambda_\text{eff}/2k> 1$ with the initial conditions near the hyperbolic fixed point $z_0=\xi_0=0$, $\phi_0=\pm\pi$ seen in Fig.~\ref{fig:phase_general}(d) ($1<\Lambda_\text{eff}/2k<2)$ and  Fig.~\ref{fig:phase_general}(e) ($ \Lambda_\text{eff}/2k>2$), whereas the (red and pink) regimes  for $\Lambda_\text{eff}/2k< -1$ with the initial conditions  near the  state $z_0=\xi_0=0$, $\phi_0=0$ in stead, which is also a hyperbolic fixed point in this range of $\Lambda_\text{eff}/2k$ in Fig.~\ref{fig:phase_general}(a) ($\Lambda_\text{eff}/2k<-2$) and in Fig.~\ref{fig:phase_general}(b) ($-1<  \Lambda_\text{eff}/2k <-2$).}
\label{fig:phase_diagram_chaos}
\end{center}
\end{figure}
 From experimental perspectives, this can be achieved by increasing the scattering length $a_{11}$ of the bilateral condensates or  decreasing the inter-species scattering length $a_{12}$, thus resulting in relatively large tunneling energy $k$.
 In Fig.~\ref{fig:chaos1}, we consider the scattering lengths giving  $\Lambda_{\text{eff}}/2k >2$
 shown in Fig.~\ref{fig:phase_diagram_chaos} for the system to illustrate the chaotic behavior.
The initial conditions are chosen near the solutions $z_0=\xi_0=0$, $\phi_0=\pm\pi$, a hyperbolic fixed point, which is a unstable point by changing $z$, realized from the effective potential in Fig.~\ref{fig:phaseportrait} and also in Table \ref{tb:limit}, but  is a stable point by changing the phase $\phi$  instead.
We solve the time-dependent GP equations using the initial condition $\psi_2(x,0)$ with $\xi (0)=0$ and $\dot \xi (0)=0$. The bilateral condensate wave function $\psi_1(x,0)$ is constructed by spatial wave functions $\psi_L $ and $\psi_R$ with the initial population imbalance $z(0)=0.09 $ and the initial relative phase difference $\phi(0)=\pi$.
%
The dynamics of $z$ is shown to run between the Josephson oscillations and the running phase MQST in Fig.~\ref{fig:chaos1}(b). With the same initial conditions for $z(0), \phi(0), \xi (0)$ and $\dot\xi (0)$, the numerical results of coupled pendulum equations  (\ref{ap3_z}), (\ref{ap3_phi}), and (\ref{ap3_xi}) also show the similar dynamics from which solutions we can compute the  so-called Lyapunov exponents for each of the dynamic variables.
A chaotic motion can be understood from the fact that, for two initial nearby initial variables $ q^{(II)}(0)\rightarrow q^{(I)}(0)$, their difference grows exponentially in time $t$, obeying
\begin{align}
&|q^{(II)}(t)-q^{(I)}(t)|=|q^{(II)}(0)-q^{(I)}(0)|e^{\lambda t}\nonumber,
\end{align}
where the rate $\lambda$ is the Lyapunov exponent obtained from
\begin{align} \label{lyap}
&\lambda = \lim_{t\rightarrow \infty}\;\lim_{{q}^{(II)}(0)\rightarrow q^{(I)}(0)} \frac{1}{t} \ln{\left|\frac{q^{(II)}(t)-q^{(I)}(t)}{q^{(II)}(0)-q^{(I)}(0)}\right|}.
\end{align}
There are four Lyapunov exponents $\lambda_{z}, \lambda_{\phi}, \lambda_{\xi}$, and $\lambda_{\dot\xi}$ shown in Fig.~\ref{fig:chaos1}(c).  The chaotic dynamics can be recognized when at least one of them is a nonzero positive value at large time, and it is found $\lambda_z >0$ in our choice of the initial conditions and scattering lengths.
\begin{figure}[t]
\begin{center}
\includegraphics[width=1\linewidth]{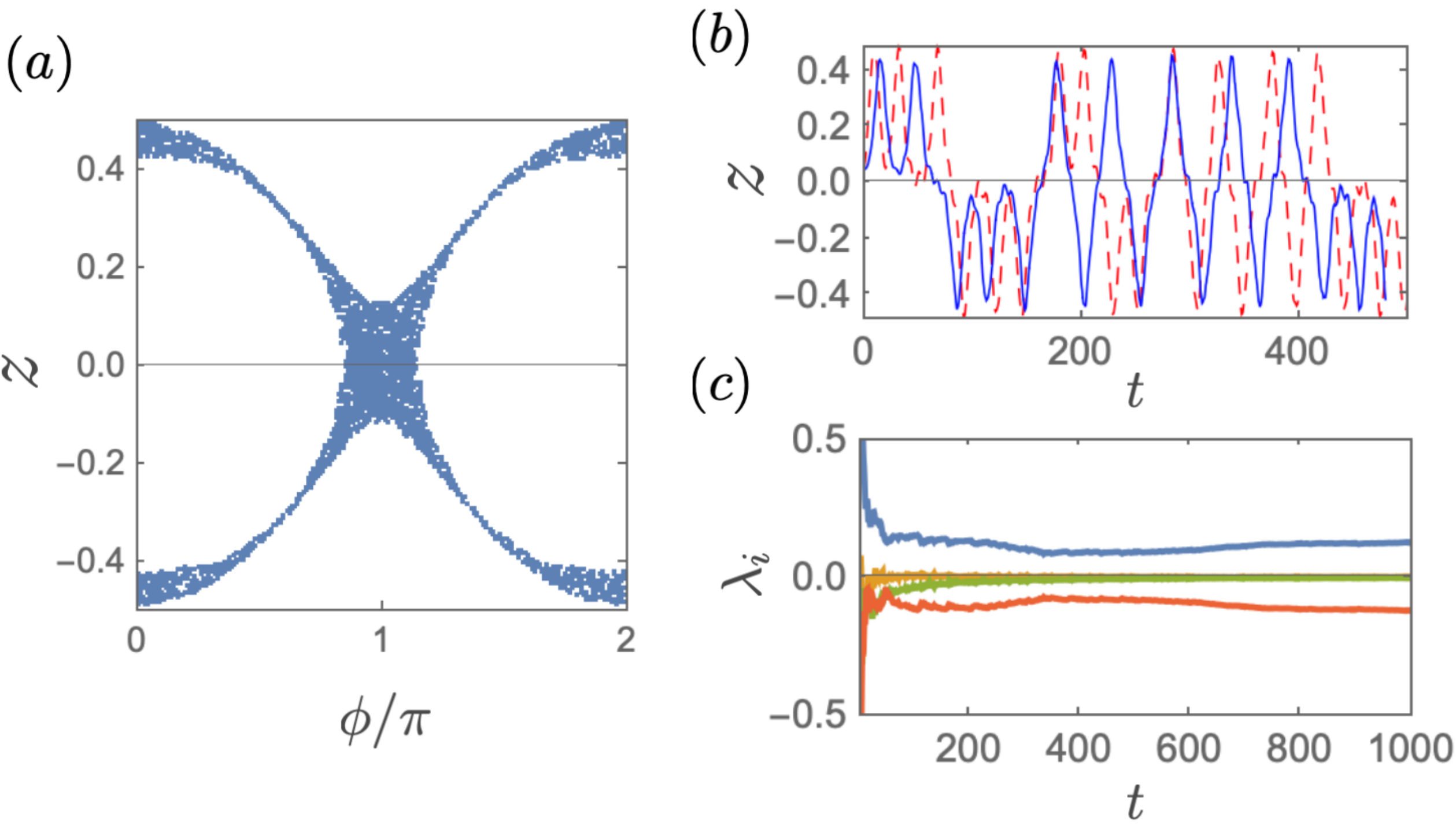}
\caption{(a) The phase portrait for the parameters: $N_1=500,\,N_2=1000,\,a_{11}=137a_{0},\, a_{22}=50a_{0}$, and $a_{12}=85a_{0}$ that result the effective parameters $\Lambda=7.1,\,\eta=2.45$, and $k=0.0345$, (b) shows chaotic oscillation with initial conditions $z(0)=0.09,\phi(0)=\pi,\,\xi(0)=\dot{\xi}(0)=0$. The red dashed line corresponds to the linearized equation Eqs.  (\ref{ap3_z}), (\ref{ap3_phi}), and (\ref{ap3_xi}) and blue line corresponds to the simulation results of real-time GP equations. (c) The corresponding Lyapunov exponents  obtained from (\ref{lyap}) are $\lambda_z$, $\lambda_{\dot\xi}$, $\lambda_{\xi}$, $\lambda_\phi$ from top to bottom.}
\label{fig:chaos1}
\end{center}
\end{figure}

In order to analytically understand  the chaotic dynamic, we start from the  CP dynamics with  Eqs.  (\ref{ap3_z})--(\ref{ap3_xi}), and substitute the solution of $\xi(t)$
in (\ref{lso12a}) and (\ref{Cxi}) into (\ref{Hz3}) where again the full dynamics of the system reduces to that of $(z,\phi)$ only.
Then the corresponding  Hamiltonian  splits into two parts
\begin{align}
H= H_z +  H_I (t),
\end{align}
where $H_z $ is the Hamiltonian to account for the dynamics of the slowly varying mode given effectively by
\begin{align}
H_z=\frac{\Lambda_\text{eff}}{2}z^2-2k\sqrt{1-z^2}\cos{\phi}
\end{align}
In the case of relatively small difference between the values of $\omega_-$ and $\omega_+$,
the dynamics  of the fast varying mode $\omega_+$ can be treated as the time dependent perturbation contributing to the interaction Hamiltonian  as
\begin{align}
H_I (t)\equiv \Delta \bar E(t) z(t) =\eta^2\omega_\xi^{-2}z(0) \cos{(\omega_+t)}z(t).
\end{align}
Nevertheless  in Refs.~ \cite{Abdullaev2000,Lee2001,Jiang2014,Tomkovic2017}, this time-dependent perturbation is implemented by introducing an external driving force.
Here, assuming $H_I \ll H_z$, one can write the solution of $z$ as
 $z=\zeta+\tilde{\zeta}$
where $\zeta$ is the solution of the unperturbed equation
\begin{align}
\ddot{\zeta}-(\Lambda_\text{eff}H_z-4k^2)\zeta+\frac{\Lambda_\text{eff}^2}{2}\zeta^3=0,
\label{unpt}
\end{align}
and $\tilde{\zeta}$ is the first-order correction in $z$ due to the perturbation $g(t)$ satisfying the equation %
\begin{align}
&\ddot{\tilde{\zeta}}-(\Lambda_\text{eff}H_z-4k^2)\tilde{\zeta}+\frac{3}{2}
\Lambda_\text{eff}{\zeta}^2\tilde{\zeta}=g(t),
\label{separatrix}\\
&g(t)=\Delta \bar{E}(t)\,  H_z +H_I \, \Lambda_\text{eff}\, \zeta-\frac{3}{2}\Lambda_\text{eff}\, \Delta \bar E(t) \, \zeta^2.
\end{align}
\begin{figure}[t]
\begin{center}
\includegraphics[width=1\linewidth]{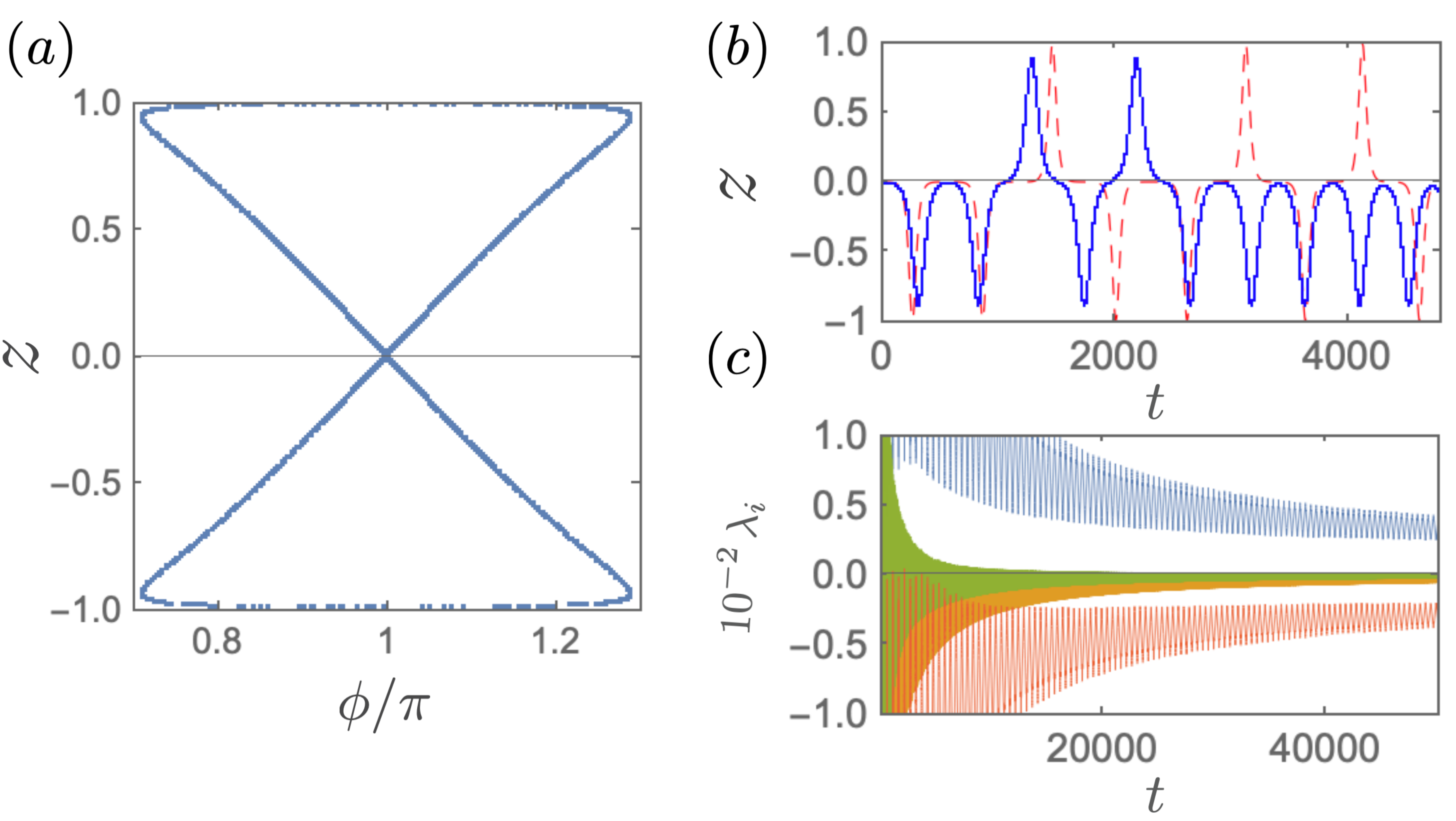}
\caption{(a) The phase portrait for the parameters: $N_1=500,\,N_2=1000,\,a_{11}=77a_{0},\, a_{22}=50a_{0}$, and $a_{12}=77a_{0}$ that result the effective parameters $\Lambda=4.8,\,\eta=2.18$, and $k=0.019$, (b) shows chaotic oscillation with initial conditions $z(0)\simeq0,\phi(0)\simeq\pi,\,\xi(0)=\dot{\xi}(0)=0$. The red dashed line corresponds to the linearized equation Eqs.  (\ref{ap3_z}), (\ref{ap3_phi}), and (\ref{ap3_xi}) and blue line corresponds to the simulation results of real-time GP equations. (c) The corresponding Lyapunov exponents  obtained from (\ref{lyap}) are $\lambda_z$, $\lambda_{\dot\xi}$, $\lambda_{\xi}$, $\lambda_\phi$ from top to bottom.}
\label{fig:chaos2}
\end{center}
\end{figure}

When $\Lambda_\text{eff}H_z-4k^2>0$, Eq. (\ref{unpt}) presents a homoclinic solution with the  double-well potential shown in Fig.~\ref{fig:phaseportrait} for $\Lambda_{\text{eff}}/2 k >2$ and  in Fig.~\ref{fig:phase_pi} for $1<\Lambda_{\text{eff}}/2 k <2$ when $H_z=2k$.
The homoclinic solution might start from a particular initial condition, and the system will drive $z$ rolling down toward one of the potential minima, and then climbing up the potential hill of  the state with $z=0$, a fixed point, just with zero velocity. The solution $\zeta$ reads as
\begin{align}
&\zeta(t)=2\sqrt{(\Lambda_\text{eff}-2 k)/\Lambda_\text{eff}^2} \,\sech(\sqrt{\Lambda_\text{eff}-2k}\,t+\mathcal{D}), \\
&\mathcal{D}=\text{arcsech}{\left(\frac{{z}(0)}{2\sqrt{(\Lambda_\text{eff}-2 k)/\Lambda_{\text{eff}}^2}}\right)}.
\end{align}
This homoclinic solution is also a separatrix, which is a boundary between the Josephson oscillations and the running phase MQST for $\Lambda_{\text{eff}}/2 k >2$, and also  the Josephson oscillations and the $\pi$ phase MQST for $1<\Lambda_{\text{eff}}/2 k <2$.
One can then construct the Melnikov function from the homogeneous solution of (\ref{separatrix}), which is $\tilde{\zeta}=\dot {\zeta}$, and the function $g(t)$ as
\begin{align} \label{M}
M(0)=\int_{-\infty}^{\infty}{dt\dot {{\zeta}}}(t)g(t)\, .
\end{align}
The existence of  zero of the Melnikov function shows the chaos \cite{Abdullaev2000,Lee2001}. Substituting Eq.~(\ref{separatrix}) into (\ref{M})  above turns out to be
\begin{align}
M(0)=&-\frac{2\pi\omega_+}{\Lambda_\text{eff}}\sqrt{\Lambda_\text{eff}-2k}\Delta \bar{E}(0)\nonumber\\
&\times \left[
\frac{H_z}{\sqrt{\Lambda_\text{eff}-2 k}}-\frac{1}{3}\sqrt{\Lambda_\text{eff}-2 k}\,\left(1+\nu^2\right)
\right]\nonumber\\
&\times \sech{\left(\frac{\pi\nu}{2}\right)}\sin{(\mathcal{D}\nu)} \, ,
\end{align}
where $\nu=\omega_+/\sqrt{\Lambda_\text{eff}-2 k}$. For the case of the fixed point $z_0= 0$, $\phi_0= \pi$, the value of $\mathcal{D} \rightarrow \infty$ means that with such highly rapidly oscillations of the sine function, any finite $\omega_{+}$ including the frequency given by the fast varying mode will make the Melnikov function zero.
Thus, the chaos occurs as shown in Fig.~\ref{fig:chaos1} for $\Lambda_{\text{eff}}/2k >2$. Similar chaos occurs when $1<\Lambda_{\text{eff}}/2k<2$, and is shown in Fig.~\ref{fig:chaos2}.
In this case, the dynamics of $z$ runs between the Josephson oscillations and the $\pi$ mode MQST instead, where its behavior can also be analyzed using the Melnikov Homoclinic method discussed above. Moreover, the chaos may appear also in $\Lambda_{\text{eff}}/2k<-1$ near the hyperbolic fixed point $z_0=0, \phi_0=0$ (see Table \ref{tb:limit}) in the regime of the scattering lengths shown in Fig.~\ref{fig:phase_diagram_chaos}.
Notice that although
the semiclassical (mean-field) GP
equation can reliably signal the onset of the chaotic motion,
it might fail to provide the details of the dynamics on the short timescales right after
entering the chaotic regime, which can be
dominated by quantum many-body effects. The exploration of chaos beyond the mean field approximation deserves further studies
~\cite{ber,Kelly2019}.

\section{Conclusions}\label{sec6}

In summary, we have proposed a new setting with binary BECs in a single-well trap potential to probe the dynamics of collective atomic motion.
In this setting,  $^{85}\text{Rb}$ atoms  $|2,-2\rangle$  and $^{87}\text{Rb}$ atoms $|1,-1\rangle$  are considered with tunable scattering lengths via Feshbach resonances so that the ground-state wave function for two types of the condensates are spatially immiscible shown in Fig.\ref{fig:phasediagram0}.
As such, the condensate of atoms for one of the hyperfine states centered at  the potential minimum can be effectively treated as a potential barrier between  bilateral condensates formed by atoms in the other hyperfine state.
In the case of small wave-function overlap of bilateral condensates, one can parametrize their spatial part of the wave functions in the two-mode approximation  together with time-dependent population of  atoms and the phase of each of the wave functions.
Besides, the wave function of the condensate in the middle is approximated  by an ansatz of the Gaussian wave function.  The
full system can be reduced to the dynamics of the imbalance population of  atoms in bilateral condensates $z$, as well as the relative phase difference $\phi$ between two wave functions   together with  the time-dependent displacement of the central condensate $\xi$.
For small wavefunction overlap of bilateral condensates shown in  Fig.~\ref{fig:phase_diagram_chaos}, all sorts of the regular trajectories, moving about the stable states, in Refs.~\cite{Smerzi1997,Raghavan1999} can be reproduced. Moreover, the numerical results given by solving the   equations of $z,\,\phi$, and $\xi$  are in close agreement with the solutions of the full time-dependent GP equations.
Nevertheless, with an increase in wave-function overlap also shown in Fig.~\ref{fig:phase_diagram_chaos}, we study the possibility of the appearance of the chaotic oscillations driven by the time-dependent displacement of the central condensate.
The  application of the Melnikov  approach with the homoclinic solutions of the   $z,\,\phi$, and $\xi$ equations successfully predicts the existence of the chaos, which are further justified from solving the full time-dependent  GP equations.
All of the findings in this work deserve further experimental investigations using advanced techniques for manipulation of atomic condensates.

\begin{acknowledgements}
This work was supported in part by the
Ministry of Science and Technology, Taiwan.
\end{acknowledgements}

\section{Appendix}
Section II has discussed a variational approach to the dynamics of binary BEC system.
In this appendix we provide more detailed derivations and approximations to arrive at the equations of the  CP dynamics given in  (\ref{ap3_z})--(\ref{ap3_xi}).
Substituting the ansatz of the ground-state wave function (\ref{psi2}) and (\ref{psi1}) into the Lagrangian (\ref{Lag}), and  carrying out the integration over space, the effective Lagrangian then becomes a functional of the time-dependent variables $\alpha$, $\beta$, $\xi$, $w$, $z$, and $\phi$ as
\begin{widetext}
\begin{align} \label{lagran_1d}
L_\text{1D}=&L_0-\frac{N_1}{2}\left(-z\dot\phi+\Delta E z+\frac{\Lambda}{2}z^2-2k_0\sqrt{1-z^2}\cos\phi\right)\nonumber\\
&-N_2\Bigg\{(\dot\alpha\xi+\dot\beta\xi^2+\frac{\dot\beta}{2}w^2)+\frac{1}{2}\left[\frac{1}{2w^2}
+2\beta^2w^2+(\alpha+2\beta\xi)^2\right]+\frac{1}{4}(2\xi^2+w^2)+\frac{g_{22}N_2}{2\sqrt{2\pi}w}
\Bigg\}\nonumber\\
&-\frac{g_{12}N_1N_2}{2\sqrt{\pi}w}\int dx \left[(\psi_L^2+\psi_R^2)+z(\psi_L^2-\psi_R^2)+2\sqrt{1-z^2}\cos\phi\psi_L\psi_R\right]\exp[-\frac{(x-\xi)^2}{w^2}]\;,
\end{align}
\end{widetext}
where the effective parameters are  tunneling energy $k_0$ (\ref{k0}), difference of energies between the wells $\Delta E$  (\ref{DE}) and self-interaction energy $\Lambda$ (\ref{lambda}).
Using Lagrangian equations for the parameters $\alpha$, $\beta$, $\xi$, $w$, $z$, and $\phi$,
we first obtain
\begin{align}
&\alpha=\dot\xi-2\xi\beta \quad \text{and}\quad\beta=\frac{\dot{w}}{2w}.\label{beta}
\end{align}
Then, after inserting Eqs.~(\ref{beta}) into (\ref{lagran_1d}), the equations of motion for population imbalance and relative phase between bilateral condensates become
\begin{align}
&\dot z+\left(2k_0-\frac{2g_{12}N_2}{\sqrt{\pi}w}\int dx \psi_L\psi_R e^{-\frac{(x-\xi)^2}{w^2}}\right)\nonumber\\&\hspace{4cm}\times\sqrt{1-z^2}\sin\phi=0\label{z},\\
&\dot \phi-\Delta E-\Lambda z\nonumber\\
&-\left(2k_0-\frac{2g_{12}N_2}{\sqrt{\pi}w}\int dx \psi_L\psi_Re^{-\frac{(x-\xi)^2}{w^2}}\right)\frac{z}{\sqrt{1-z^2}}\cos\phi\nonumber\\
&-\frac{g_{12}N_2}{\sqrt{\pi}w}\int dx (\psi_L^2-\psi_R^2)e^{-\frac{(x-\xi)^2}{w^2}}=0\label{phi}.
\end{align}
For the center condensate, the equations of motion of $\xi$ and $w$
are obtained as
\begin{align}
&\ddot{\xi}+\xi+\frac{g_{12}N_1}{\sqrt{\pi}w^3}\int dx\left[(\psi_L^2+\psi_R^2)+z(\psi_L^2-\psi_R^2)\right.\nonumber\\
& \left.\qquad +2\sqrt{1-z^2}\cos\phi\psi_L\psi_R\right](x-\xi)e^{-\frac{(x-\xi)^2}{w^2}}=0,\label{xi}
\end{align}
\begin{align}
&\ddot{w}+w-\frac{1}{w^3}-\frac{g_{22}N_2}{\sqrt{2\pi}w^2}+\frac{g_{12}N_1}{\sqrt{\pi}w^2}\int dx\Big[(\psi_L^2+\psi_R^2)\nonumber\\
&\quad+z(\psi_L^2-\psi_R^2)+2\sqrt{1-z^2}\cos\phi\psi_L\psi_R\Big]\nonumber\\
&\qquad\times\left[\frac{2(x-\xi)^2}{w^2}-1\right]e^{-\frac{(x-\xi)^2}{w^2}}=0.\label{sigma}
\end{align}
It can be understood that the presence of bilateral condensates contribute a time-dependent deformation for the central condensate by coupling to population imbalance $z(t)$ in the integrand of Eqs.~(\ref{xi}) and (\ref{sigma}).
We introduce Gaussian functions
(see Fig.~\ref{fig:wavefun})
%
\begin{align}
&\psi_{L}(x)=\left(\frac{1}{\pi\lambda^2}\right)^{1/4}e^{-\frac{(x+\zeta)^2}{2\lambda^2}},\\
&\psi_{R}(x)=\left(\frac{1}{\pi\lambda^2}\right)^{1/4}e^{-\frac{(x-\zeta)^2}{2\lambda^2}},
\label{twogauss}
\end{align}
and substitute them into Eqs. (\ref{z})--(\ref{sigma}). In our cases, the displacement $\xi$ and the width defined as $w=\sigma_0+\sigma$ with  $\sigma_0$ determined initially by solving time-independent GP equation for finding the ground state solution and $\sigma$ driven by $\xi$ , satisfy the conditions
\begin{align}
\xi \ll \sqrt{\lambda^2+\sigma_0^2}\quad\text{and}\quad \sigma \ll \sqrt{\lambda^2+\sigma_0^2}\,.
\end{align}
\begin{widetext}
In the case of $\lambda \sim \sigma_0$, we have $\xi \ll  \sqrt{\lambda^2+\sigma_0^2}\sim\sigma_0$ and $\sigma \ll  \sqrt{\lambda^2+\sigma_0^2}\sim\sigma_0$. Back to Eqs.~(\ref{z})--(\ref{sigma}),  it is allowed to expand the equations in terms of small $\sigma/\sigma_0$ and $\xi/\sigma_0$ as
\begin{align}
&\dot z+\left[2k_0-\frac{2g_{12}N_2}{\sqrt{\pi}\sigma_0}\left(1-\frac{\sigma}{\sigma_0}+\mathcal{O}\left(\frac{\sigma^2}{\sigma_0^2}\right)\right)\int dx \psi_L\psi_R e^{-\frac{(x-\xi)^2}{\sigma_0^2}(1-2\sigma/\sigma_0)}\right]\sqrt{1-z^2}\sin\phi=0\;,
\end{align}
where the integral above can be further expanded as
\begin{align}
&\left[1-\frac{\sigma}{\sigma_0}+\mathcal{O}\left(\frac{\sigma^2}{\sigma_0^2}\right)\right]
\int dx \psi_L\psi_R e^{{\left[-\left(\frac{x}{\sigma_0}\right)^2+2\left(\frac{x\xi}{\sigma_0^2}\right)+\mathcal{O}\left(\frac{\xi^2}{\sigma_0^2}\right)\right]
\left(1-2\frac{\sigma}{\sigma_0}+\mathcal{O}\left(\frac{\sigma^2}{\sigma_0^2}\right)\right)}} \nonumber\\[7pt]
&= \int \,dx \psi_L\psi_R e^{-{x^2}/{\sigma_0^2}}\left[1-\frac{\sigma}{\sigma_0}\left(1-2\frac{x^2}{\sigma_0^2}\right)\right]+\int dx \psi_L\psi_R e^{-{x^2}/{\sigma_0^2}}\left(\frac{2 x\xi}{\sigma_0^2}\right)+\mathcal{O}\left(\frac{\xi^2}{\sigma_0^2},\;\frac{\sigma^2}{\sigma_0^2}\right)\, .
\end{align}
The term of order $\xi/\sigma_0$ vanishes due to the odd function in the integrand. Therefore Eq.~(\ref{z}) can be further simplified by keeping the terms of order $\sigma/\sigma_0$ as
\begin{align}
&\dot z+\left[2k_0-\frac{2g_{12}N_2}{\sqrt{\pi}\sigma_0}\int{dx\psi_L\psi_Re^{-x^2/\sigma_0^2}}+\frac{2g_{12}N_2}{\sqrt{\pi}\sigma_0}\int dx\psi_L\psi_Re^{-x^2/\sigma_0^2}\left(1-2\frac{x^2}{\sigma_0^2}\right)\left(\frac{\sigma}{\sigma_0}\right)\right]
\sqrt{1-z^2}\sin\phi=0,\label{ap_z}
\end{align}
and reduces to (\ref{ap2_z}) with the definition of the constants in (\ref{k}) and (\ref{kapp}).
Following the same procedure to approximate Eq. (\ref{phi}), we have
\begin{align}
&\dot \phi-\Delta E-\Lambda z-\left[2k_0-\frac{2g_{12}N_2}{\sqrt{\pi}\sigma_0}\int{dx\psi_L\psi_Re^{-x^2/\sigma_0^2}}+\frac{2g_{12}N_2}{\sqrt{\pi}\sigma_0}\int dx\psi_L\psi_Re^{-x^2/\sigma_0^2}\left(1-2\frac{x^2}{\sigma_0^2}\right)\left(\frac{\sigma}{\sigma_0}\right)\right]\nonumber\\
&\hspace{2.5cm}\times \frac{z}{\sqrt{1-z^2}}\cos\phi- \frac{2g_{12}N_2}{\sqrt{\pi}\sigma_0^3}\int dx (\psi_L^2-\psi_R^2)\,x\,e^{-\frac{x^2}{\sigma_0^2}} \xi=0 \; , \label{ap_phi}
\end{align}
which leads to (\ref{ap2_phi}).

Now we turn to linearize Eq. (\ref{xi}), which is
\begin{align}
&\ddot{\xi} +\xi + \frac{g_{12}N_1}{\sqrt{\pi}\sigma_0^3}\left[1-3\frac{\sigma}{\sigma_0}+\mathcal{O}\left(\frac{\sigma^2}{\sigma_0^2}\right)\right]\int dx \left[(\psi_L^2+\psi_R^2)+z(\psi_L^2-\psi_R^2)+2\sqrt{1-z^2}\cos\phi\psi_L\psi_R\right]\nonumber\\
&\qquad\times(x-\xi)e^{-{x^2}/{\sigma_0^2}}\left[1-\frac{x^2}{\sigma_0^2}+2\left(\frac{x}{\sigma_0}\right)\left(\frac{\xi}{\sigma_0}\right)+2\left(\frac{x^2}{\sigma_0^2}\right)\left(\frac{\sigma}{\sigma_0}\right)+\mathcal{O}\left(\frac{\xi^2}{\sigma_0^2},\,\frac{\sigma^2}{\sigma_0^2}\right)\right]=0.\nonumber\\
\end{align}
 Considering the vanishing of the integral due to the odd function in the integrand, we conclude
\begin{align}
&\ddot{\xi}+\left[1+\frac{g_{12}N_1}{\sqrt{\pi}\sigma_0^3}\int dx (\psi_L^2+\psi_R^2)\left(2\frac{x^2}{\sigma_0^2}-1\right)e^{-x^2/\sigma_0^2}\right]\xi+\left[\frac{g_{12}N_1}{\sqrt{\pi}\sigma_0^3}\int dx(\psi_L^2-\psi_R^2)xe^{-x^2/\sigma_0^2}\right]z\nonumber\\
&\hspace{6cm}+\left[\frac{g_{12}N_1}{\sqrt{\pi}\sigma_0^3}\int dx (\psi_L^2-\psi_R^2)\left(2\frac{x^3}{\sigma_0^3}-3\frac{x}{\sigma_0}\right)e^{-x^2/\sigma_0^2}\right]\sigma=0,\label{ap_xi}
\end{align}
that gives (\ref{ap2_xi}).
Finally, we can linearize Eq.~(\ref{sigma}) as
\begin{align}
&\ddot{\sigma}+\left[1+\frac{3}{\sigma_0}+\frac{g_{22}N_2}{\sqrt{2\pi}\sigma_0^3}+\frac{g_{12}N_1}{\sqrt{\pi}\sigma_0^3}\int_{-\infty}^{\infty}dx(\psi_L^2+\psi_R^2)\left(2-10\frac{x^2}{\sigma_0^2}+4\frac{x^4}{\sigma_0^4}\right)e^{-x^2/\sigma_0^2}\right]\sigma\nonumber\\
&\hspace{6cm}-\left[\frac{2g_{12}N_1}{\sqrt{\pi}\sigma_0^3}\int_{-\infty}^{\infty}dx(\psi_L^2-\psi_R^2)\left(3\frac{x}{\sigma_0}-2\frac{x^3}{\sigma_0^3}\right)e^{-x^2/\sigma_0^2}\right]\xi=0,\label{ap_sigma}
\end{align}
giving (\ref{ap2_sigma}).
\end{widetext}

\end{document}